\documentclass[11pt,a4paper]{article}
\usepackage{epsf,epsfig,psfrag,graphics,color}
\usepackage{amsmath}
\usepackage{geometry}
\usepackage{amsfonts}
\usepackage{mathrsfs}
\usepackage{amssymb}
\usepackage{bm}
\usepackage{bbm}
\usepackage{wasysym}
\usepackage{cite}
\newcommand{\beq}{\begin{equation}}
\newcommand{\eeq}{\end{equation}}
\newcommand{\bqa}{\begin{eqnarray}}
\newcommand{\eqa}{\end{eqnarray}}

\def\slash#1{#1 \hskip-0.5em /}

\begin{document}

\title{$B_c$ Exclusive Decays to Charmonium and a Light Meson at Next-to-Leading Order Accuracy}

\author{Cong-Feng
Qiao$^{a}$\footnote{E-mail:qiaocf@ucas.ac.cn}, Peng
Sun$^{b,c}$\footnote{E-mail:sunp@pku.edu.cn}, Deshan
Yang$^{a}$\footnote{E-mail:yangds@ucas.ac.cn}, and Rui-Lin
Zhu$^{a}$\footnote{E-mail:zhuruilin09@mails.ucas.ac.cn}}
\date{}
\maketitle

\vspace*{-0.5cm}
\begin{center}
$^a$ Department of Physics, University of
Chinese Academy of Sciences \\
YuQuan Road 19A, Beijing 100049, China\\ $^b$ Center for
High-Energy Physics, Peking University, Beijing 100871, China\\
$^c$ Nuclear Science Division, LBNL, Berkeley, CA 94720, USA
\end{center}

\vspace*{1cm}
\begin{abstract}
In this paper the next-to-leading order (NLO) corrections to $B_c$
meson exclusive decays to S-wave charmonia and light pseudoscalar or
vector mesons, i.e. $\pi$, $K$, $\rho$, and $K^*$, are performed
within non-relativistic (NR) QCD approach. The non-factorizable
contribution is included, which is absent in traditional naive
factorization (NF). And the theoretical uncertainties for their
branching ratios are reduced compared with that of direct tree level
calculation. Numerical results show that NLO QCD corrections
markedly enhance the branching ratio with a K factor of 1.75 for
$B_{c}^{\pm}\to \eta_{c} \pi^{\pm}$ and  1.31 for $B_{c}^{\pm}\to
J/\psi \pi^{\pm}$. In order to investigate the asymptotic behavior,
the analytic form is obtained in the heavy quark limit, i.e. $m_b
\to \infty$. We note that annihilation topologies contribute trivia
in this limit, and the corrections at leading order  in $z= m_c/m_b$
expansion come from form factors and hard spectator interactions. At
last, some related phenomenologies are also discussed.

\begin{description}
\item[PACS numbers] 12.38.Bx, 12.39.St, 13.20.He
\end{description}

\end{abstract}

\clearpage

\thispagestyle{empty}

\newpage
\setcounter{page}{1}

\section{Introduction}

$B_c$ and its excited states construct the unique meson family
containing two different kinds of heavy flavor. The studies on
production and decay of $B_c$ can shed light on the understanding of
the strong interaction in such a unique system. In contrast to other
bottom mesons just embodying one heavy flavor which can be produced
remarkably through the $e^+ e^-$ and $ep$ collisions, the cross
section of $B_c$ is suppressed owing to the associated production of
two additional heavy quarks $c$ and $\bar{b}$\cite{HQP}. Thus
massive $B_c$ events can but refer to the hadron colliders.

After the first discovery of $B_c$ was reported by the CDF
collaboration at Tevatron in 1998 \cite{Bc:1998}, there are
continuous measurements of its mass in different detectors via two
different channels: $B_{c}^{\pm}\to J/\psi \ell^{\pm} v_\ell$
\cite{CDF:2006,D0:2009} and $B_{c}^{\pm}\to J/\psi \pi^{\pm}$
\cite{CDF:2008,D0:2008}. Especially for the latter exclusive
two-body decay, it takes advantage of a large trigger efficiency.
Using this channel, the LHCb collaboration have measured the $B_c$
mass with $6273.7\pm 1.3(stat)\pm1.6(sys) \mathrm{MeV}/c^2$ recently
\cite{LHCb:2012-9}.  However, the exact value of branching ratio for
either $B_{c}^{\pm}\to J/\psi \ell^{\pm} v_\ell$ or $B_{c}^{\pm}\to
J/\psi \pi^{\pm}$ has not been measured yet. And more channels
should be involved  to investigate the intrinsic properties of
$B_c$.  Up to now, the LHCb collaboration have successfully observed
more channels beyond the two kinds in question.  And they have
measured the new channels $B_{c}^{+}\to J/\psi \pi^+\pi^-\pi^+$
\cite{LHCb:2012},  $B_{c}^{\pm}\to J/\psi K^{\pm}$
\cite{Aaij:2013vcx}, $B_{c}^{\pm}\to \Psi(2S) \pi^{\pm}$
\cite{LHCb:2013}, and $B_{c}^{\pm}\to J/\psi D_s^+$
\cite{Aaij:2013gia} for the first time.  The study of decay
properties of $B_c$ from a multitude of processes can help us to
understand the quark flavor mixing and provide precision
determination of the CKM matrix parameters. Besides, according to
Refs~\cite{BCevent,BCevent2,LHCb:2010}, the cross section of $B_c$
is expected to $\sim40 nb$ at the $pp$ center-of-mass energy
$\sqrt{s}=14\mathrm{TeV}$. That means around $10^{10}$ $B_c$ meson
per year can be anticipated at LHC. Thus one should expect a more
variety of decay channels of $B_c$ can be measured in the upcoming
experiment.

Theoretically, the  exclusive two-body dacay of the bottom meson is
studied within the frame of the naive factorization, potential
model, pQCD method and QCD factorization in the heavy quark limit.
Along with the technique for the QCD factorization for the exclusive
hard processes, such as $\pi$ electromagnetic form-factors at the
large momentum transfer and $B$ meson decays to two light mesons,
many theorists believe that the QCD factorization for $B_{c}^{-}\to
J/\psi \pi^{-}$ holds in the heavy quark limit generally. However,
there is no complete or consistent predictions based at NLO in
$\alpha_{s}$ so far.

Since $B_c^-$ contains two kinds of heavy quark, namely $b$ and $c$
quarks, the heavy quark limit may be realized in NRQCD approach.
Therein one lets $m_b, m_c\to \infty$ and keeps the ratio $z\equiv
m_c/m_b$ fixed. Then the decay amplitude of $B_c^-\to
J/\psi(\eta_{c}) \pi^-$ is conjectured to be factorized
\begin{eqnarray}
{\cal A}(B_{c}^{-}\to J/\psi(\eta_{c})\pi^{-})\sim \Psi_{c\bar c}
(0)\Psi_{b\bar c}(0) \int_{0}^{1}dx
T_{H}(x,\mu)\phi_{\pi}(x,\mu)+{\cal O}(1/m_{b})+ {\cal
O}(v^2)\,.\label{eq:NR}
\end{eqnarray}
Here $\Psi_{c\bar c}(0)$ and $\Psi_{b\bar c}(0)$ denote the
Schr\"odinger wave functions at origin of $J/\psi(\eta_{c})$ and
$B_c^-$, respectively; $T_{H}(x,\mu)$ is the perturbatively
calculable hard kernel; and $\phi_{\pi}(x,\mu)$ is the Pion's
light-cone distribution amplitude (LCDA).

The rough arguments of the validity of the above factorization are:
1) the energetic Pion ejected from the heavy quark system, cannot
sense the surrounded soft and collinear gluons, due to the
``color-transparency" at the leading order of heavy quark expansion,
the hadronization of the collinear quark pair into a Pion is totally
described by the leading twist LCDA of Pion, as the case in $B\to
\pi\pi$; 2) the charm-quark in $B_c$ needs a large momentum transfer
(typically $q^2\sim m_b m_c\sim 6 {\rm GeV}^2$) to speed up for
catching another energetic charm-quark from the $b\to c$ weak
transition to form a quarkonia. This large momentum transfer
guarantees the necessary condition for the implementation of the
perturbative QCD in this process, i.e. the transition from $B_c$  to
$J/\psi (\eta_c)$ at the large recoil can be described the
hard-gluon exchange, and the hadronization is to be described by the
non-relativistic wave functions (at the origin) of $B_c$ and $J/\psi
( \eta_c)$, as what done in many NRQCD factorization for the
exclusive quarkonia processes.

In this paper, we will adopt the factorization formula (\ref{eq:NR}) to
 calculate $B_c\to J/\psi (\eta_c) \pi$ to the next-to-leading order of
 strong coupling $\alpha_s$. In our calculation, we do find that all the
 low-energy divergences, including soft, collinear and Coulomb divergences,
 are either cancelled with each other (for the soft interactions), or separated
 with each other to be absorbed into the LCDA and the wave functions. Thus,
 our work can be treated as a proof for the factorization formula (\ref{eq:NR}) at one-loop level.

The following sections are organized as follows: in
Sect.~\ref{sect:Ham} we present a brief overview of the effective
weak Hamiltonian; and in Sect.~\ref{sect:NR} we present the detailed
computation in the NR factorization scheme, we also deliver the asymptotic behavior in the limit $z=m_c/m_b\to 0$;
 in Sect.~\ref{sect:Num} we implement our results to make
some phenomenological predictions for the branching ratios of
various $B_c$ two-body decays to a S-wave quarkonium and a light
meson,  and some detailed
 discussions are also presented; at last we conclude in Sect.~\ref{sect:con}.

\section{The theoretical frame\label{sect:Ham}}
In the Standard Model (SM), $B_{c}^{-}\to J/\psi \pi^{-}$ occurs
through $W$-mediated charge current process. However, since
$m_{W}>>m_{b,c},~\Lambda_{\rm QCD}$, the large logarithm arise in
the higher order strong interaction corrections. Thus, the
RG-improved perturbation theory must be resorted. In the community
of $B$ physics, this turns to be the effective weak Hamiltonian
method. The effective weak Hamiltonian governing $B_{c}^{-}\to
J/\psi \pi^{-}$ is
\begin{eqnarray}
{\cal H}_{\rm
eff}=\frac{G_{F}}{\sqrt{2}}V_{ud}^{*}V_{cb}\left(C_1(\mu) Q_{1}(\mu)
+C_{2}(\mu)Q_{2}(\mu)\right)\,,
\end{eqnarray}
with $G_{F}$ being the Fermi constants, $V_{ud}$ and $V_{cb}$ the
Cabibbo-Kobayashi-Maskawa (CKM) matrix-elements, $C_{1,2}(\mu)$ the
perturbatively calculable Wilson coefficients, and $Q_{1,2}(\mu)$
the effective four-quark operators
\begin{subequations}
\begin{eqnarray}
Q_{1}&=&\bar d_{\alpha}\gamma^{\mu}(1-\gamma_{5})u_{\alpha}\bar
c_{\beta}\gamma_{\mu}
(1-\gamma_{5})b_{\beta}\,,\\
Q_{2}&=&\bar d_{\alpha}\gamma^{\mu}(1-\gamma_{5})u_{\beta}\bar
c_{\beta}\gamma_{\mu} (1-\gamma_{5})b_{\alpha}\,,
\end{eqnarray}
\end{subequations}
where $\alpha,\beta$ are color indices and the summation convention
over repeated indices are understood. For the conveniences of our
later calculations, we will adopt another operator basis, i.e.
\begin{subequations}
\begin{eqnarray}
Q_{0}&=&\bar d_{\alpha}\gamma^{\mu}(1-\gamma_{5})u_{\alpha}\bar
c_{\beta}\gamma_{\mu}
(1-\gamma_{5})b_{\beta}\,,\\
Q_{8}&=&\bar
d_{\alpha}T^{A}_{\alpha\beta}\gamma^{\mu}(1-\gamma_{5})u_{\beta}\bar
c_{\rho}T_{\rho\lambda}^{A}\gamma_{\mu}(1-\gamma_{5})b_{\lambda}\,,
\end{eqnarray}
\end{subequations}
where $T^{A}$s are the generators of the fundamental representation
for ${\rm SU}_{\rm C}(3)$. Applying the Fierz rearrangement relation
\begin{eqnarray}
T^{A}_{\alpha\beta}T^{A}_{\rho\lambda}=-\frac{1}{6}\delta_{\alpha\beta}
\delta_{\rho\lambda}+\frac{1}{2}\delta_{\alpha\lambda}\delta_{\rho\beta}\,,
\end{eqnarray}
we have immediately
\begin{eqnarray}
Q_{0}=Q_{1}\,,~~Q_{8}=-\frac{1}{6}Q_{1}+\frac{1}{2}Q_{2}\,.
\end{eqnarray}
Consequently, for the Wilson coefficients, we have
\begin{eqnarray}
C_{0}=C_{1}+C_{2}/3\,,~~C_{8}=2 C_{2}\,.
\end{eqnarray}
Then, the decay amplitude of $B_{c}^{-}\to J/\psi(\eta_{c})\pi^{-}$
can be written as
\begin{eqnarray}
{\cal A}(B_{c}^{-}\to J/\psi(\eta_{c})\pi^{-})&=&\left\langle J/\psi
(\eta_{c})\pi^{-}\left\vert{\cal H}_{\rm eff}\right\vert
B_{c}^{-}\right
\rangle\nonumber\\
&=&\frac{G_{F}}{\sqrt{2}}V_{ud}^{*}V_{cb}\left(C_0(\mu) \langle
Q_{0}(\mu)\rangle+C_{8}(\mu)\langle Q_{8}(\mu)\rangle\right)\,.
\end{eqnarray}

\section{The non-relativistic approach\label{sect:NR}}

Systematically, the non-relativistic QCD effective theory provides
an rigorous factorization formalism for the annihilation and
production of heavy quarkonia\cite{Bodwin:1995}. In this framework,
the heavy quarkonium's production comes from two steps: a Fock state
such as $|q\bar{q}\rangle$, $|q\bar{q}g\rangle$ produced at
short-distance by a large momentum transfer process, followed by it
binding to quarkonium  at long-distance.

In the process of $B_{c}^{-}\to J/\psi(\eta_{c})\pi^{-}$ ,
 all the non-perturbative blinding effects are attributed to three
factors: Pion decay constant and the Schr\"odinger wave functions at
origin of $J/\psi(\eta_{c})$ and $B_c$. While the hard kernel can be
calculated perturbatively.

\subsection{LO}

\begin{figure}[h]
 \centering
\includegraphics[width=0.80\textwidth]{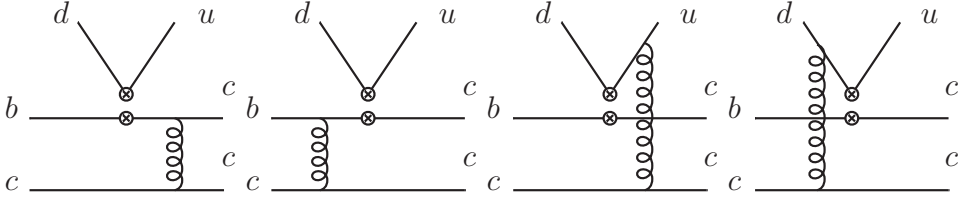}
\caption{\small The quark-level Feynman diagrams at leading order
for $B_{c}\to J/\psi(\eta_{c})\pi$. The 4-vertex ``$\otimes
~~\otimes$'' denotes the insertion of a 4-fermion operator $Q_i$.
}\label{Fig:tree}
\end{figure}

The possible quark-level topologies for $B_{c}\to
J/\psi(\eta_{c})\pi$ are portrayed in Figure~\ref{Fig:tree}, where
we assign momentum xP to the u-quark  and   (1-x)P to the d-quark in
the emitted Pion. The former two in Figure~\ref{Fig:tree} contribute
to $\langle Q_{0}\rangle$, and the others contribute to $\langle
Q_{8}\rangle$. It is completely perturbatively calculable sector.
Associated with non-perturbative parameters : Pion decay constant
and the Schr\"odinger wave functions at origin of $J/\psi(\eta_{c})$
and $B_c$, we have the tree-level $\langle Q_{i}\rangle$, leaving
the momentum fraction x unintegrated
\begin{eqnarray}
\langle Q_{0} (\eta_c)\rangle_x &=& \frac{8 \sqrt{2} \pi  f_{\pi }
\psi_{\eta_c}(0) \psi_{B_c}(0)\phi_\pi(x) C_A C_F \alpha _s
\sqrt{m_b+m_c} \left(m_b+3 m_c\right) \left(2 m_b m_c+3 m_b^2+3
m_c^2\right)}{m_c^{3/2} N_c
\left(m_b-m_c\right){}^3}\,,\nonumber\\
\langle Q_{8} (\eta_c)\rangle_x &=&\frac{2 \sqrt{2} \pi  f_{\pi }
\psi_{\eta_c}(0) \psi_{B_c}(0)\phi_\pi(x) C_A C_F \alpha _s
\sqrt{m_b+m_c} \left(m_b+3 m_c\right){}^2 \left(x m_c-(x-1)
m_b\right)}{m_c^{3/2} N_c^2 \left(m_c-m_b\right) \left((x-1) m_b +(3
x-2) m_c\right) \left(x m_b+(3 x-1)
m_c\right)},~~~~~\label{eq:Qi_etac_x}
\end{eqnarray}
where more detail about $f_{\pi }$, $\psi_{\eta_c}(0)$,
$\psi_{B_c}(0)$, and Pion light cone distribution amplitude
$\phi_\pi(x,\mu)$ can be found in Appendix~A. Note that higher twist
contribution comes from twist-4. Referring to  $J/\psi$, the
corresponding matrix elements are
\begin{eqnarray}
\langle Q_{0} (J/\psi)\rangle_x &=& -\frac{64 \sqrt{2} \pi
 f_{\pi } \psi_{J/\psi}(0) \psi_{B_c}(0)\phi_\pi(x)
 P_{B_c}\!\!\cdot\!\varepsilon^*_\Psi
C_A C_F \alpha _s \left(m_b+m_c\right){}^{5/2}}{m_c^{1/2} N_c
\left(m_b-m_c\right){}^4}\,,\nonumber\\
\langle Q_{8} (J/\psi)\rangle_x &=& -\frac{8 \sqrt{2} \pi f_{\pi }
\psi_{J/\psi}(0)  \psi_{B_c}(0)\phi_\pi(x)
P_{B_c}\!\!\cdot\!\varepsilon^*_\Psi C_A C_F \alpha _s
\left(m_b+m_c\right){}^{1/2}}{m_c^{1/2} N_c^2 \left
(m_b-m_c\right){}^2 \left((x-1) m_b+(3 x-2) m_c\right) \left(x
m_b+(3 x-1)
m_c\right)}\nonumber\\
&& \times  \left(3 (2 x-1) m_b m_c+(x-1) m_b^2+(9 x-4)
m_c^2\right)\,.\label{eq:Qi_jpsi_x}
\end{eqnarray}

Note that $\langle Q_{8}\rangle$ in Equation~(\ref{eq:Qi_etac_x})
and (\ref{eq:Qi_jpsi_x}) is not symmetrical when exchange x with
$\bar{x}=1-x$, because of the non-factorizable contribution from
axial vector current  which brings in an anti-symmetrical part.
However, the anti-symmetrical part can be easily proved to be
insignificant. We define the function $V(x)$ to collect the
contributions from axial vector current, and it satisfies
$V(\bar{x})=-V(x)$. Considering the symmetrical Pion LCDA, i.e.
$\phi_\pi(\bar{x})=\phi_\pi(x)$,  we can get the result
\begin{eqnarray}
\int_0^1V(x)\phi_\pi(x)dx &=&
-\int_1^0V(\bar{x})\phi_\pi(\bar{x})dx=-\int_0^1V(x)\phi_\pi(x)dx=0.
\end{eqnarray}

Employing the asymptotic LCDA $\phi_\pi(x,\mu\to\infty)=6x\bar{x}$,
we can obtain the integrated matrix elements $\langle Q_{i}\rangle$
\begin{eqnarray}
\langle Q_{0} (\eta_c)\rangle &=& \frac{8 \sqrt{2} \pi  f_{\pi }
\psi_{\eta_c}(0) \psi_{B_c}(0) C_A C_F \alpha _s \sqrt{m_b+m_c}
\left(m_b+3 m_c\right) \left(2 m_b m_c+3 m_b^2+3
m_c^2\right)}{m_c^{3/2} N_c
\left(m_b-m_c\right){}^3}\,,\nonumber\\
\langle Q_{8} (\eta_c)\rangle &=& \frac{6 \sqrt{2} \pi  f_{\pi }
\psi_{\eta_c}(0) \psi_{B_c}(0) C_A C_F \alpha _s \sqrt{m_b+m_c}}
{m_c^{3/2} N_c^2 \left(m_b-m_c\right) \left(m_b+3 m_c\right)}
\times [2 m_b m_c (\ln (m_b+2 m_c)\nonumber\\
&& -\ln (m_c)+2)+m_c^2 (4\ln (m_b+2 m_c)-4 \ln (m_c)+3)+m_b^2]\,,\nonumber\\
\langle Q_{0} (\Psi)\rangle &=& -\frac{64 \sqrt{2} \pi f_{\pi }
\psi_{J/\psi}(0)  \psi_{B_c}(0) P_{B_c}\!\!\cdot\!\varepsilon^*_\Psi
C_A C_F \alpha _s \left(m_b+m_c\right){}^{5/2}}{m_c^{1/2} N_c
\left(m_b-m_c\right){}^4}\,,\nonumber\\
\langle Q_{8} (\Psi)\rangle &=&- \frac{24 \sqrt{2} \pi  f_{\pi }
\psi_{J/\psi}(0)  \psi_{B_c}(0) P_{B_c}\!\!\cdot\!\varepsilon^*_\Psi
C_A C_F \alpha _s \left(m_b+m_c\right){}^{1/2}\times  [2 m_b m_c
(\ln \left(m_b+2 m_c\right)}
{m_c^{1/2} N_c^2 \left(m_b-m_c\right) \left(m_b+3 m_c\right){}^3}\nonumber\\
&&  -\ln \left(m_c\right)+2)+m_c^2 \left(4 \ln \left(m_b+2
m_c\right)-4 \ln\left(m_c\right)+3\right)+m_b^2] \,.
\end{eqnarray}

Rather than the traditional formalisms in
Refs.~\cite{Buchalla:1996,Beneke:2000}, herein we
 extracted hard kernels $T_{i}$ from Wilson coefficients separately.  They can be calculated perturbatively order by
 order.
\begin{equation}
   {\cal A}(B_{c}^{-}\to J/\psi(\eta_{c})\pi^{-})=
   \frac{G_{F}}{\sqrt{2}}V_{ud}^{*}V_{cb}\left(C_0(\mu)  T_{f,0}M_{f}+C_0(\mu)  T_{nf,0}M_{nf}
   +C_{8}(\mu)T_{nf,8}M_{nf}\right)\label{eq:expand},
\end{equation}
\begin{equation}
  T_{f,i}(\mu)=\sum_{k=0}^\infty (\frac{\alpha_s}{4\pi})^k
  T^{(k)}_{f,i}(\mu)\,,~~~T_{nf,i}(\mu)=\sum_{k=0}^\infty (\frac{\alpha_s}{4\pi})^k
  T^{(k)}_{nf,i}(\mu)\,,
\end{equation}
where $T_{f}$ means factorizable hard kernel, $T_{nf}$ means
non-factorizable hard kernel. And the
 Wilson coefficients $C_{i}$ are
\begin{eqnarray}
C_{0}=\frac{2}{3}C_{+}+\frac{1}{3}C_{-}\,,~~C_{8}=C_{+}-C_{-}\,,
\end{eqnarray}
where
\begin{eqnarray}
C_{\pm}=\left[\frac{\alpha_s(M_W)}{\alpha_s(\mu)}\right]^{
\frac{\gamma_\pm} {2\beta_0}}\,,~~\gamma_\pm=\pm 6\frac{N_c\mp
1}{N_c}\,,~~\beta_0=\frac{11N_c-2n_f}{3}\,.
\end{eqnarray}

Fixing $M_{f}(\eta_c)=\langle Q_{0} (\eta_c)\rangle$,
$M_{f}(J/\psi)=\langle Q_{0} (\Psi)\rangle$, $M_{nf}(\eta_c)=\langle
Q_{8} (\eta_c)\rangle$ and $M_{nf}(J/\psi)=\langle Q_{8}
(\Psi)\rangle$, we can extract the leading order hard kernel
$T^{(0)}_{i}$
\begin{eqnarray}
T^{(0)}_{f,0}(\eta_c)=T^{(0)}_{f,0}(J/\psi)=1\,,~~T^{(0)}_{nf,0}(\eta_c)
=T^{(0)}_{nf,0}(J/\psi)=0,~~T^{(0)}_{nf,8}(\eta_c)=T^{(0)}_{nf,8}(J/\psi)=1\label{eq:extract}\,.
\end{eqnarray}

\subsection{NLO}

Now we pay more attention  to the corrections  at next-to-leading
order. The  one loop diagrams  for $B_{c}\to J/\psi(\eta_{c})\pi$
are classified into Figures~\ref{Fig:factorizaiton},~\ref{Fig:o1}
and~\ref{Fig:o8}. Where Fig.~\ref{Fig:factorizaiton} lays out the
one loop factorizable diagrams while non-factorizable diagrams are
shown in Figs~\ref{Fig:o1} and~\ref{Fig:o8}.  To regularize the
Ultra-Violet and Infre-Red divergences we use dimensional
regularization scheme, but relative velocity regularization scheme
for Coulomb divergence. The renormalization constants are listed in
Appendix~\ref{renormal}. In our calculation, the Mathematical
package FeynArts\cite{Feynarts} was used to generate the Feynman
diagrams, FeynCalc\cite{Feyncalc} to deal with the amplitudes, and
LoopTools\cite{Looptools} to calculate the one-loop integrals.  The
practicable $\gamma_5$-scheme is adopted in D dimensional
computation\cite{Korner,Qiao:2011}.
\subsubsection{$T^{(1)}_{0}$}

\begin{figure}[h]
 \centering
\includegraphics[width=0.80\textwidth]{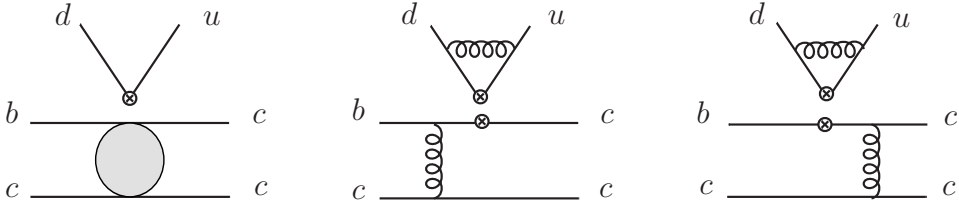}
\caption{\small One loop factorizable diagrams contribute to
$\langle Q_{0}\rangle$. The bubble in the first diagram expresses
all one loop form factor diagrams, which  are displayed  in
Refs~\cite{Bell:2007,Sun:2011}.}\label{Fig:factorizaiton}
\end{figure}
\begin{figure}[h]
 \centering
\includegraphics[width=0.80\textwidth]{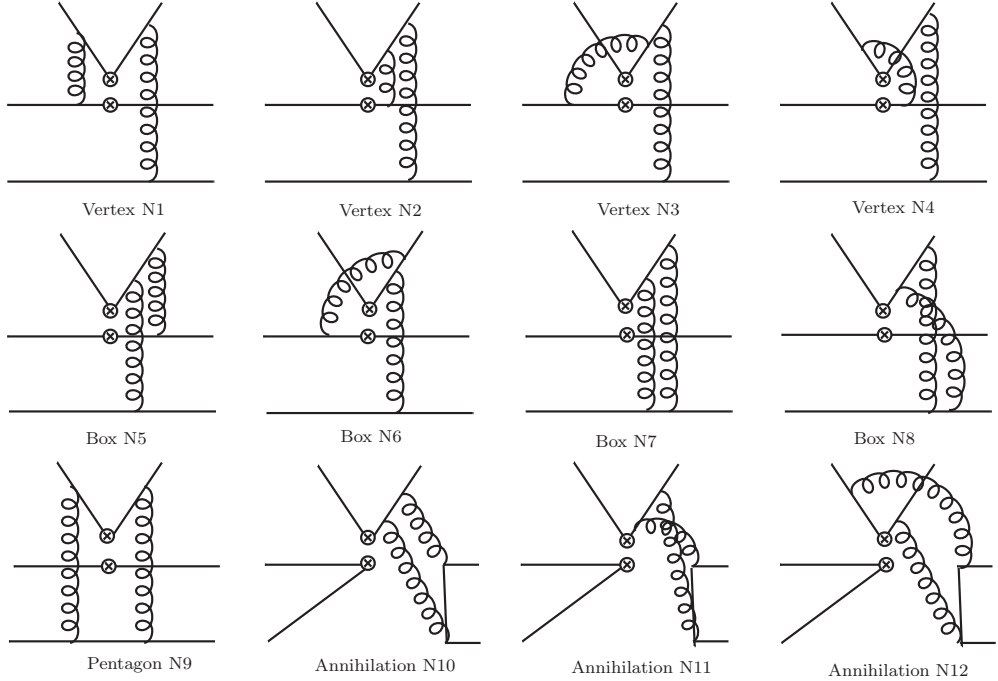}
\caption{\small Twelve of  twenty-four one loop non-factorizable
diagrams contribute to $\langle Q_{0}\rangle$. The other twelve
partners can be obtained by interchanging u and d
quarks.}\label{Fig:o1}
\end{figure}

The one loop diagrams contribute to $\langle Q_{0}\rangle$ can be
distributed into two sets: factorizable(see
Fig.~\ref{Fig:factorizaiton}) and non-factorizable diagrams(see
Fig.~\ref{Fig:o1}). And the NLO $B_c$-to-S-wave-charmonium form
factors  have been calculated in Refs
\cite{Bell:2007,Bell:2008,Sun:2011,Qiao:2012}. For the rear two in
Fig.~\ref{Fig:factorizaiton}, their UV divergence can be canceled by
external field counter terms.

We analyzed the factorizable part at first. In NF, $\langle
Q_{8}\rangle$ vanishes, and
\begin{eqnarray}
\langle J/\psi(\eta_{c})\pi^{-}\vert Q_{0}\vert
B_{c}^{-}\rangle\approx \langle J/\psi(\eta_{c})\vert \bar c
\gamma^{\mu}(1-\gamma_{5})b\vert B_{c}^{-}\rangle\langle
\pi^{-}\vert \bar d\gamma_{\mu}(1-\gamma_{5})u\vert 0\rangle\,,
\end{eqnarray}
i.e. $\langle Q_{0}\rangle$ is proportional to the product of the
Pion decay constant and $B_{c}^{-}\to J/\psi (\eta_{c})$ transition
form-factor. Conventionally, we adopt the following
parameterizations for the decay constants and $B_{c}^{-}\to J/\psi
(\eta_{c})$ transition form-factors
\begin{eqnarray}
\langle \eta_{c}(p)\vert\bar c\gamma_{\mu}\gamma_{5} c\vert 0\rangle
&=&-if_{\eta_{c}}p_{\mu}\,,
\\
\langle \pi^{-}(p^{\prime})\vert\bar d\gamma_{\mu}\gamma_{5} u\vert
0\rangle &=&-if_{\pi}p^{\prime}_{\mu}\,,
\\
\langle B_{c}^{-}(P)\vert\bar c\gamma_{\mu}\gamma_{5} b\vert
0\rangle &=&-if_{B_{c}}P_{\mu}\,,
\\
\langle J/\psi(p,\varepsilon^{*})\vert\bar c\gamma_{\mu} c\vert
0\rangle &=&-if_{J/\psi}m_{J/\psi}\varepsilon^{*}_\mu\,,
\\
\langle \eta_{c}(p)\vert \bar c \gamma^{\mu}b\vert
B_{c}^{-}(P)\rangle
&=&f_{+}(q^{2})\left[P^{\mu}+p^{\mu}-\frac{m_{B_{c}}^{2}-
m_{\eta_{c}}^{2}}{q^{2}}q^{\mu}\right]+f_{0}(q^{2})
\frac{m_{B_{c}}^{2}-m_{\eta_{c}}^{2}}{q^{2}}q^{\mu}\,,\\
\langle \eta_{c}(p)\vert \bar c \gamma^{\mu}\gamma_{5}b\vert
B_{c}^{-}(P)\rangle
&=&0\,,\\
\langle J/\psi(p,\varepsilon^{*})\vert \bar c \gamma^{\mu}b\vert
B_{c}^{-}(P)\rangle &=&\frac{2 i
V(q^{2})}{m_{B_{c}}+m_{J/\psi}}\epsilon^{\mu\nu\rho\sigma}
\varepsilon_{\nu}^{*}p_{\rho}P_{\sigma}\,,
\\
\langle J/\psi(p,\varepsilon^{*})\vert \bar c
\gamma^{\mu}\gamma_{5}b\vert B_{c}^{-}(P)\rangle &=&2
m_{J/\psi}A_{0}(q^{2})\frac{\varepsilon^{*}\cdot
q}{q^{2}}q^{\mu}+(m_{B_{c}}+m_{J/\psi})A_{1}(q^{2})
\left[\varepsilon^{*\mu}-\frac{\varepsilon^{*}\cdot q}{q^{2}}
q^{\mu}\right]\nonumber\\
&&-A_{2}(q^{2})\frac{\varepsilon^{*}\cdot
q}{m_{B_{c}}+m_{J/\psi}}\left[
P^{\mu}+p^{\mu}-\frac{m_{B_{c}}^{2}-m_{J/\psi}^{2}}{q^{2}}q^{\mu}\right]\,,
\end{eqnarray}
here we define momentum transfer $q=P-p$ and $\epsilon^{0123}=-1$.
Note that $f_{0}(0)=f_{+}(0)$.

The tree-level form factors can be obtained easily. They read

\bqa f_+^{LO}(q^{2})=\frac{8 \sqrt{2} C_A C_F \pi  \sqrt{z+1}
\left(-\frac{q^2}{m_b^2}+3 z^2+2 z+3\right) \alpha _s \psi(0)_{B_c}
\psi(0)_{\eta_c}}{\left(\frac{q^2}{m_b^2}-(z-1)^2\right)^2 z^{3/2}
m_b^3 N_c}\; , \eqa
\bqa f_0^{LO}(q^{2})=\frac{8 \sqrt{2} C_A C_F \pi  \sqrt{z+1}
\left(9 z^3+9 z^2+11 z-\frac{q^2}{m_b^2} (5 z+3)+3\right) \alpha _s
\psi(0)_{B_c}
\psi(0)_{\eta_c}}{\left(\frac{q^2}{m_b^2}-(z-1)^2\right)^2 z^{3/2}
(3 z+1) m_b^3 N_c}\; ,
 \eqa
\bqa V^{LO}(q^{2})=\frac{16 \sqrt{2} C_A C_F \pi  (3 z+1) \alpha _s
\psi(0)_{B_c}\psi(0)_{J/\psi}}{\left(\frac{q^2}{m_b^2}-(z-1)^2\right)^2
   \left(\frac{z}{z+1}\right)^{3/2} m_b^3 N_c}\; ,
 \eqa
\bqa A_0^{LO}(q^2)=\frac{16 \sqrt{2} C_A C_F \pi (z+1)^{5/2} \alpha
_s\psi(0)_{B_c}\psi(0)_{J/\psi}}{\left(\frac{q^2}{m_b^2}-(z-1)^2\right)^2
z^{3/2} m_b^3 N_c}\; ,
 \eqa
\bqa A_1^{LO}(q^2)=\frac{16 \sqrt{2} C_A C_F \pi \sqrt{z+1} \left(4
z^3+5 z^2+6 z-\frac{q^2}{m_b^2} (2 z+1)+1\right) \alpha _s
\psi(0)_{B_c}\psi(0)_{J/\psi}}{\left(\frac{q^2}{m_b^2}-(z-1)^2\right)^2
z^{3/2} (3 z+1) m_b^3 N_c}\; ,
 \eqa
\bqa A_2^{LO}(q^2)=\frac{16 \sqrt{2} C_A C_F \pi \sqrt{z+1} (3 z+1)
\psi(0)_{B_c}\psi(0)_{J/\psi}}{\left(\frac{q^2}{m_b^2}-(z-1)^2\right)^2
z^{3/2} m_b^3 N_c}\; ,
 \eqa
here,  $z\equiv m_c/m_b$.

When neglecting the mass of Pion, we have the factorizable
contribution in NF
\begin{eqnarray}
&&\langle \eta_{c}\pi^{-}\vert Q_{0,f}\vert B_{c}^{-}\rangle= i
f_{\pi} f_{0}(0)(m_{B_{c}}^{2}-m_{\eta_{c}}^{2})\,,\nonumber\\
&&\langle J/\psi\pi^{-}\vert Q_{0,f}\vert B_{c}^{-}\rangle= -i
f_{\pi} A_{0}(0)(m_{B_{c}}^{2}-m_{J/\psi}^{2}) \,,
\end{eqnarray}
where we have used the fact that $J/\psi$ is longitudinally
polarized so that
\begin{eqnarray}
2 m_{J/\psi} \varepsilon^{*}\cdot P=2 m_{B_{c}}\vert
\vec{p}_{c}\vert=m_{B_{c}}^{2}-m_{J/\psi}^{2}\,.\nonumber
\end{eqnarray}
Therefore we have the LO result
\begin{eqnarray}
\langle \eta_{c}\pi^{-}\vert Q_{0}\vert
B_{c}^{-}\rangle^{LO}&=&i\,\frac{8 \sqrt{2}  \pi\alpha _s C_A C_F
\sqrt{z+1} \left(9 z^3+9 z^2+11 z+3\right)
 f_{\pi}\psi(0)_{B_c}
\psi(0)_{\eta_c}}{(1-z)^3 z^{3/2} m_b N_c}\,,
\end{eqnarray}
and in the heavy quark limit $z\to 0$, we have
\begin{eqnarray}
\lim\limits_{z\to 0}\langle \eta_{c}\pi^{-}\vert Q_{0}\vert
B_{c}^{-}\rangle^{LO}&=&i\,\frac{24 \sqrt{2}  \pi\alpha _s C_A C_F
 f_{\pi}\psi(0)_{B_c}
\psi(0)_{\eta_c}}{ z^{3/2} m_b N_c}\,,
\end{eqnarray}
Actually the approximation above is not so good. Numerically, we
have
\begin{eqnarray}
\left.\frac{\lim\limits_{z\to 0}\langle \eta_{c}\pi^{-}\vert
Q_{0}\vert B_{c}^{-}\rangle^{LO}}{\langle \eta_{c}\pi^{-}\vert
Q_{0}\vert B_{c}^{-}\rangle^{LO}}\right\vert_{z=1.5/4.8}\approx
0.11\,,
\end{eqnarray}
which is essentially bad. The perturbative series expanded as
equation (\ref{eq:expand}), however, can resolve the problem.
Because the convergence of hard kernel $T_i$  is well-behaved.

Note that the complete analytic expression is too lengthy to
presented and it is possible to derive an asymptotic analytic
formula valid in phenomenological application. Thus, we present our
results in the heavy quark limit, i.e. $m_b \to \infty$.

The factorizable  hard kernel $T^{(1)}_{f,0}$ is identical to the
ratio of NLO form factor to the tree level one

\begin{eqnarray}
T^{(1)}_{f,0}
(\eta_c)=\frac{f^{(1)}_{0}(0)}{f^{(0)}_{0}(0)}&=&\frac{1}{3} (11
C_A-2 \text{n}_f) \ln
 (\frac{2\mu^2}{z m_b^2})-\frac{10\text{n}_f}{9}-\frac{1}{3} \ln
 z-\frac{2 \ln 2}{3}\nonumber\\
&&+C_F\big(\frac{1}{2} \ln ^2z+\frac{10}{3} \ln 2 \ln z-\frac{35}{6}
\ln z+\frac{2 \ln
   ^22}{3}\nonumber\\
&&+3 \ln 2+\frac{7 \pi ^2}{9}-\frac{103}{6}\big)\nonumber\\
&&+C_A\big(-\frac{1}{6} \ln ^2z-\frac{1}{3} \ln 2 \ln z-\frac{1}{3}
\ln z+\frac{\ln
   ^22}{3}\nonumber\\
&&-\frac{4 \ln 2}{3}-\frac{5 \pi ^2}{36}+\frac{73}{9}\big)\,,
\end{eqnarray}
\begin{eqnarray}
T^{(1)}_{f,0}
(\Psi)=\frac{A^{(1)}_{0}(0)}{A^{(0)}_{0}(0)}&=&\frac{1}{3} (11 C_A-2
\text{n}_f) \ln (\frac{2\mu^2}{z m_b^2})-\frac{10\text{n}_f}{9}
+C_F\big(\frac{1}{2} \ln ^2z-\frac{119}{8}\nonumber\\
&&+7 \ln 2 \ln z-\frac{21}{4} \ln z+7 \ln
^22+\frac{15 \ln 2}{4}\big)\nonumber\\
&&+C_A\big(-\frac{3}{8} \ln ^2z-\ln 2 \ln z-\frac{9}{8}
\ln z-\frac{7 \pi ^2}{24}+\frac{67}{9}\nonumber\\
&&-\frac{9 \ln ^22}{4}+\frac{3 \ln 2}{8}\big)\,.
\end{eqnarray}

\begin{table}
\centering
 \caption{\label{tab:color-o1}Color factors of the corresponding diagrams
 in Fig.~\ref{Fig:o1}. Therein the figures
from N1 to N9, contribute not only  $\langle Q_{0}\rangle$, but also
$\langle Q_{8}\rangle$.} \hspace{-2.5mm}
\begin{tabular}{|c||c|c|c|}\hline

Diagram &  N1-2,N6,N8-9 &N3-5,N7& N10-12
\\\hline color in $\langle Q_{8}\rangle$ &
$\frac{(C_A^2-2)C_F}{4}$ &$-\frac{C_F}{2}$ & 0
\\\hline color in $\langle Q_{0}\rangle$ & \multicolumn{3}{|c|}{
$\frac{C_A C_F}{2}$}
\\\cline{2-4}\hline
\end{tabular}
\end{table}

Next, we turn to the non-factorizable part. There are twenty-four
one loop non-factorizable diagrams contribute to $\langle
Q_{0}\rangle$, half of which are displayed in Fig.~\ref{Fig:o1}.
 The corresponding color factors
are summed up in Table~\ref{tab:color-o1}.  Over one hundred of
one-loop integrals  in Fig.~\ref{Fig:o1} and Fig.~\ref{Fig:o8} are
created, and they can be reduced into Master Integrals(MI) and some
two-point Passarino-Veltman  integrals\cite{Looptools}.  Our
analytic expressions of MI are in agreement with what given in
Refs~\cite{Ellis:2008,Dittmaier:2003}.

The  numerical one loop non-factorizable contribution for
$T^{(1)}_{nf,0}$ are
\begin{eqnarray}
T^{(1)}_{nf,0} (\eta_c)&=&6\ln
(\frac{m_b^2}{\mu^2})+16.75\label{NRo0}\,,
\end{eqnarray}
\begin{eqnarray}
T^{(1)}_{nf,0}(\Psi)&=&T^{(1)}_{nf,0} (\eta_c)\,.
\end{eqnarray}
And the complete  results in heavy quark limit can be found in
Appendix~C.

\subsubsection{$T^{(1)}_{8}$}
\begin{figure}[h]
 \centering
\includegraphics[width=0.80\textwidth]{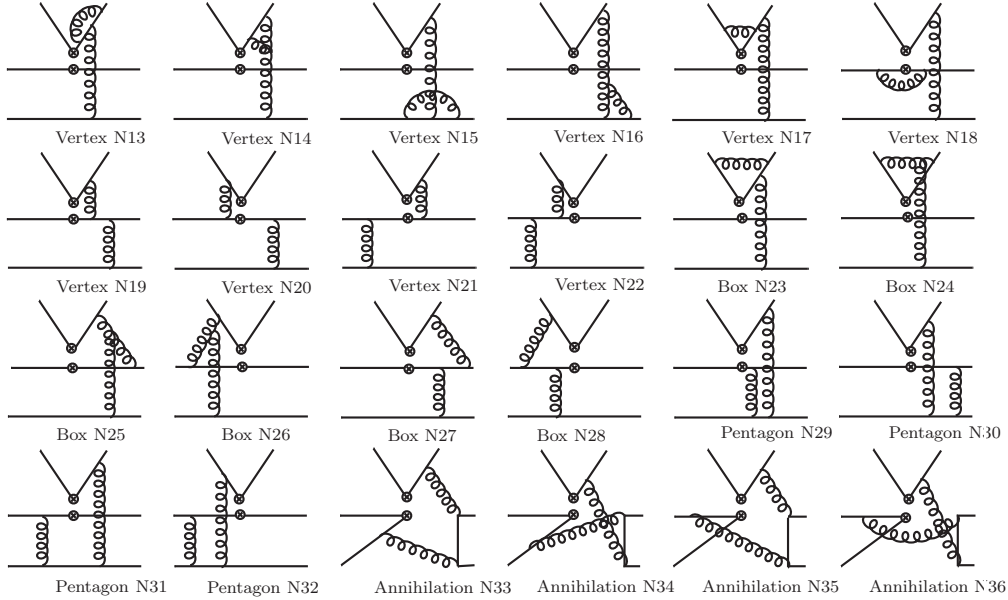}
\caption{\small Twenty-four of the sixty-five one loop
non-factorizable diagrams contribute to $\langle Q_{8}\rangle$.
Another twenty-three diagrams can be obtained by interchanging u and
d quark.  And the left eighteen come from the diagrams Vertex N1 to
Pentagon N9 in Fig.~\ref{Fig:o1} and their symmetrical
partners.}\label{Fig:o8}
\end{figure}

Here, we study the one loop non-factorizable  contributions to
$\langle Q_{8}\rangle$. Twenty-four diagrams are showed up in
Fig.~\ref{Fig:o8}, another nine diagrams are collected in
Fig.~\ref{Fig:o1} from N1 to N9, and the rest ones can be obtained
by interchanging u and d quarks. The corresponding color factors are
summed up in Table~\ref{tab:color-o1} and Table~\ref{tab:color-o8}.
\begin{table}
\centering
 \caption{\label{tab:color-o8}Color factors of the corresponding diagrams
 in Fig.~\ref{Fig:o8}. They contribute  only to $\langle Q_{8}\rangle$.}
\hspace{-1.5mm}
\begin{tabular}{|c||c|c|c|c|}\hline

Diagram &N13,N15,N17-18,N27-29,N33-36 & N14,N26 & N16,N24-25 &
N19-23,N30-32
\\\hline color in $\langle
Q_{8}\rangle$ &$\frac{2 C_A C_F}{3}$ & $\frac{iC_A^2 C_F}{4} $ &
$-\frac{iC_A^2 C_F}{4} $ & $-\frac{C_AC_F}{12}$ \\\hline
\end{tabular}
\end{table}

After integrating the fraction,  we have the corresponding
$T^{(1)}_{nf,8}$
\begin{eqnarray}
T^{(1)}_{nf,8}(\eta_c)&=&\frac{1}{3} \left(-11 C_A+2 n_f+16
N_c-6\right)\ln (\frac{m_b^2}{\mu^2})-\frac{12.48}{N_c}+(9 \ln
z+1)C_F\nonumber\\
&&-(\frac{\ln ^2z}{2}-\frac{6 \ln 2-23}{3} \ln
z+278.1)C_A-\frac{2}{9} n_f (-3
\ln z+5+3 \ln 2)\nonumber\\
&&+\frac{\ln ^2z}{6}-\frac{8(3+\ln 2)}{3} \ln z
+548.9\label{NRo8}\,,
\end{eqnarray}
\begin{eqnarray}
T^{(1)}_{nf,8}(\Psi)&=&\frac{1}{9} \left(-33 C_A+6 n_f+32
N_c-18\right)\ln (\frac{m_b^2}{\mu^2})-\frac{12.48}{N_c}+(9 \ln
z+1)C_F\nonumber\\
&&-(\frac{\ln ^2z}{2}-\frac{6 \ln 2-23}{3} \ln
z+278.1)C_A-\frac{2}{9} n_f
 (-3 \ln z+5+3 \ln 2)\nonumber\\
&&+\frac{\ln ^2z}{6}-\frac{8(3+\ln 2)}{3} \ln z+542.3\,.
\end{eqnarray}

\section{The phenomenological studies\label{sect:Num}}

The decay width  can be written as:
\begin{eqnarray}
\Gamma(B_c\to J/\psi(\eta_{c})\pi)&=&\frac{|p|}{8\pi
m_{B_c}^2}|{\cal A}(B_{c}\to J/\psi(\eta_{c})\pi)|^2,
\end{eqnarray}
here the momentum of final particle
 satisfies $|p|=(m_{B_c}^2-m_{\Psi}^2)/2m_{B_c}$ in the $B_c$ meson rest frame,
and we adopt the input parameters as below\cite{PDG:2010}
\begin{eqnarray}
&&m_c=1.4\pm0.1\mathrm{GeV},~m_b=4.9\pm0.1\mathrm{GeV},~\Lambda_{QCD}=100\mathrm{MeV},
~G_F=1.16637\times10^{-5}\mathrm{GeV}^{-2}
\,,\nonumber\\
 &&|V_{ud}^*V_{cb} |=A\lambda^2 (1 - \lambda^2/2-
\lambda^4/8),~~~n_f=3,~~~N_c=3,~~~C_F=4/3,~~~|V_{us}^*|=0.2252\,,\nonumber\\
&&|V_{cb}|=0.0406,~~~f_\pi=130.4\mathrm{MeV},~~f_\rho=216\mathrm{MeV},~~f_K=156.1\mathrm{MeV},
~~f_K^*=220\mathrm{MeV}\,,\nonumber
\end{eqnarray}
where $A=0.814$, $\lambda=0.2257$. The Schr\"{o}dinger wave function
at the origin for  $J/\psi$ is determined through its leptonic decay
width $\Gamma_{ee}^\psi=5.55\mathrm{keV}$\cite{PDG:2010}.
Numerically we can obtain  $|\psi
_{\Psi}^{LO}(0)|^2=0.0447(\mathrm{GeV})^3$ and $|\psi
_{\Psi}^{NLO}(0)|^2=0.0801(\mathrm{GeV})^3$. For that of $B_c$, we
shall determine its value to be:
$|\psi_{B_c}(0)|^2=0.1307(\mathrm{GeV})^3$, which is derived under
the Buchm\"{u}ller-Tye potential\cite{Eichten:1994}. Besides, the
one loop result for strong coupling constant is used, i.e.
\begin{eqnarray}
\alpha_s(\mu)=\frac{4\pi}{(11-\frac{2}{3}n_f)\ln(\frac{\mu^2}
              {\Lambda_{QCD}^2})}\,.\nonumber
\end{eqnarray}

\begin{figure}
\centering
\includegraphics[width=0.480\textwidth]{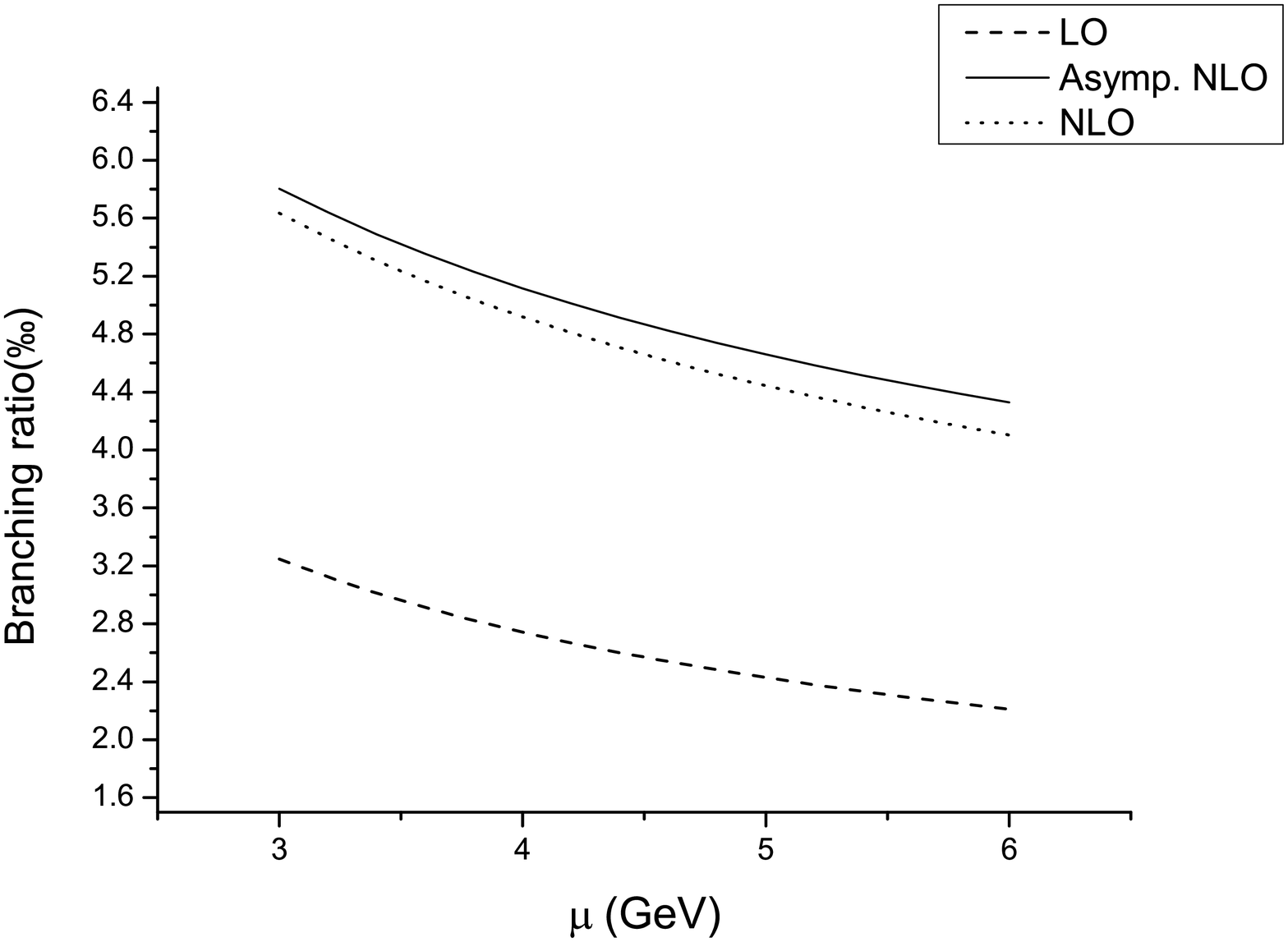}%
\hspace{5mm}
\includegraphics[width=0.480\textwidth]{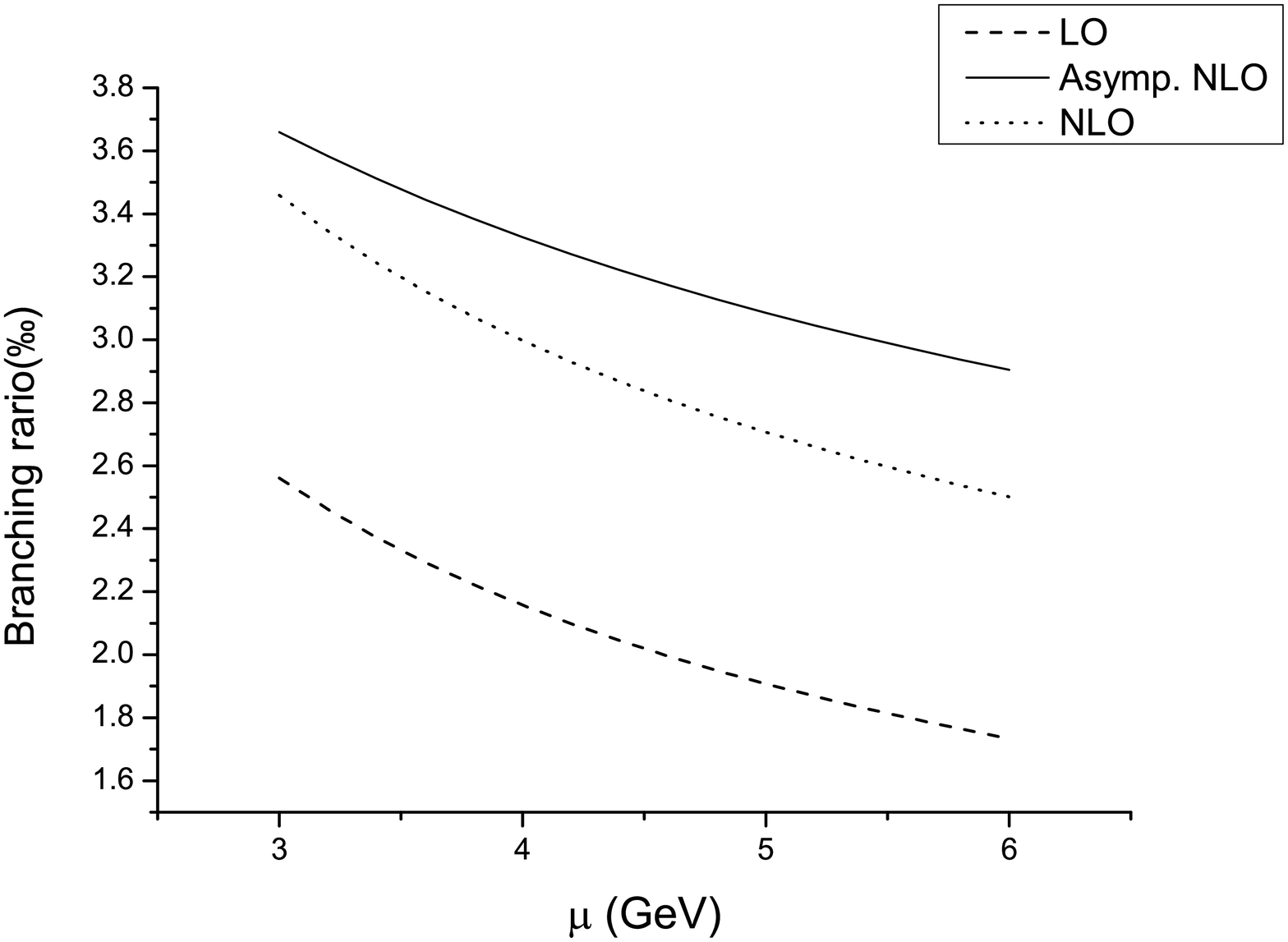}\hspace*{\fill}
\caption{\small The branching ratios of $B_{c}\to \eta_{c}
\pi$(left) and $B_{c}\to J/\psi \pi$(right) versus renormalization
scale $\mu$. Herein $m_c=1.5\mathrm{GeV},~m_b=4.8\mathrm{GeV}$, and
for the lifetime of the $B_c$ we take $\tau(B_c) =0.453$ ps. The
results of LO, Asymptotic NLO, and complete NLO  are shown.}
\label{Fig:Branchings} \vspace{-0mm}
\end{figure}

\begin{figure}
\centering
\includegraphics[width=0.480\textwidth]{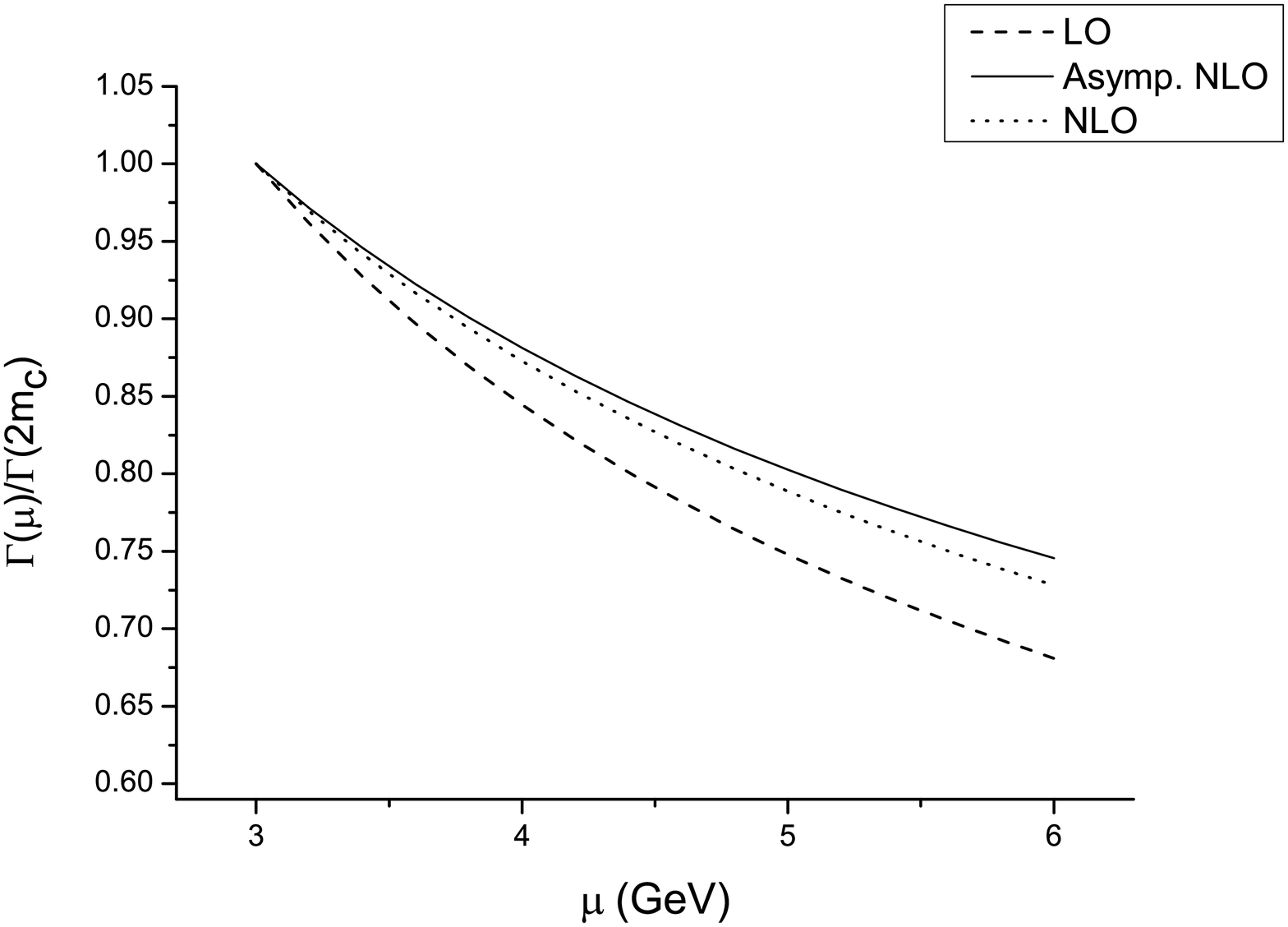}%
\includegraphics[width=0.480\textwidth]{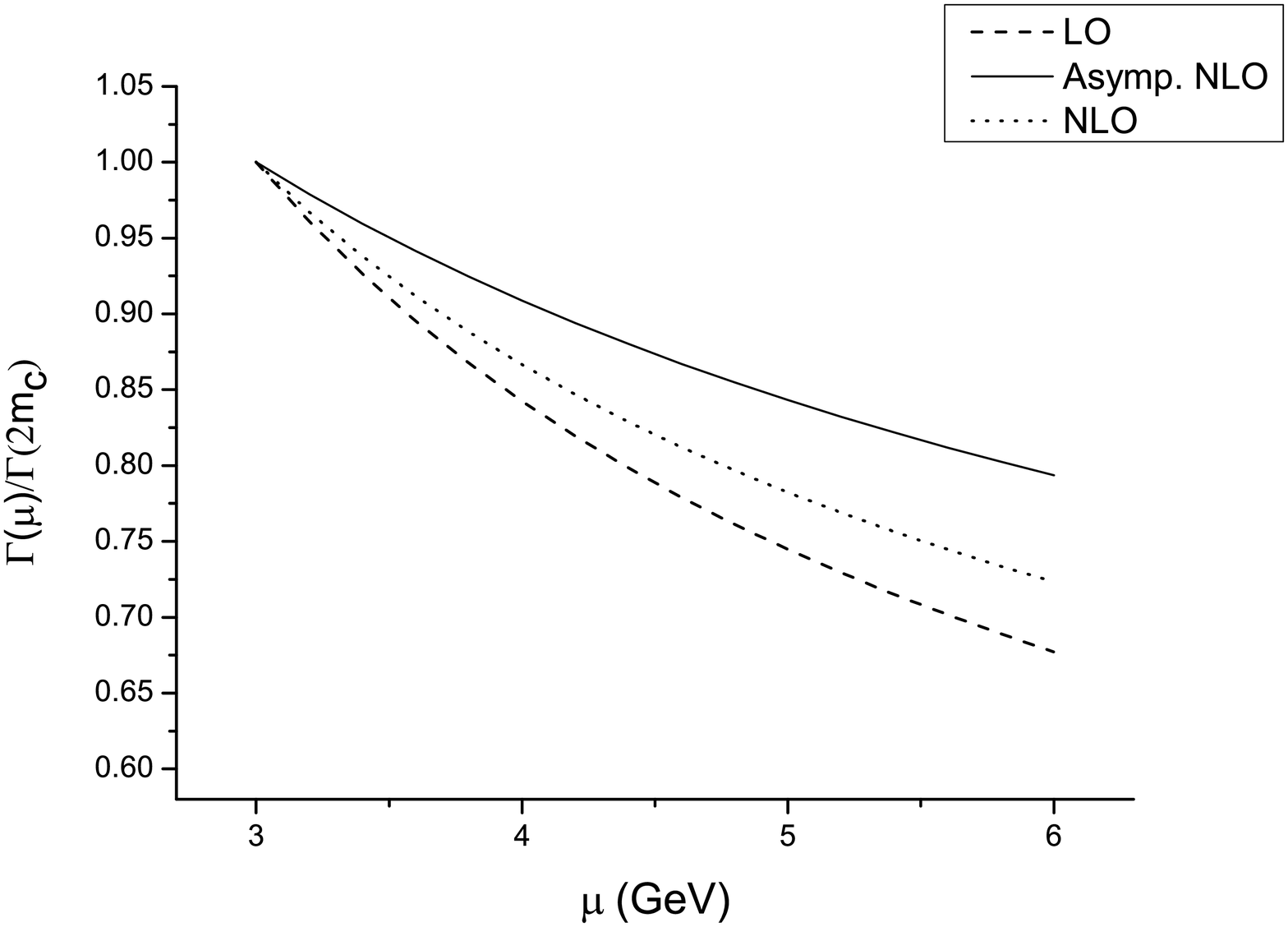}\hspace*{\fill}
\caption{\small The ratio $\Gamma(\mu)/\Gamma(2m_c)$ of $B_{c}\to
\eta_{c} \pi$(left) and $B_{c}\to J/\psi \pi$(right) versus
renormalization scale $\mu$. Herein
$m_c=1.5\mathrm{GeV},~m_b=4.8\mathrm{GeV}$. } \label{Fig:mu_depnd}
\vspace{-0mm}
\end{figure}

Within the above input parameters, we can obtain the decay width of
$B_c$ decays to S-wave charmonium and Pion at NLO accuracy. In
practise, the renormalization scale $\mu$ may run from $2m_c$ to
$m_{b}$, and the  $\mu$ dependence  of branching ratio is shown in
Figure~\ref{Fig:Branchings}. Therein, we plot both kinds of NLO
results: one letting $m_c/m_b\rightarrow~0$ in heavy-quark-limit;
the other fixing the ratio $m_c/m_b$ to its physical value. The
first one is valid in leading $m_c/m_b$, while the latter summed to
all orders of $m_c/m_b$. It turned out the leading order
approximation in $m_c/m_b$ expansion, namely Asymptotic NLO result,
account for more than $85\%$ of the complete NLO result. That means
it is enough for us to use this  simple and analytic expression for
phenomenological studies in place of complicated NLO expression. The
NLO corrections can reduce the uncertainty, which is explicitly
exhibited  in Figure~\ref{Fig:mu_depnd}.

\begin{figure}
\centering
\includegraphics[width=0.480\textwidth]{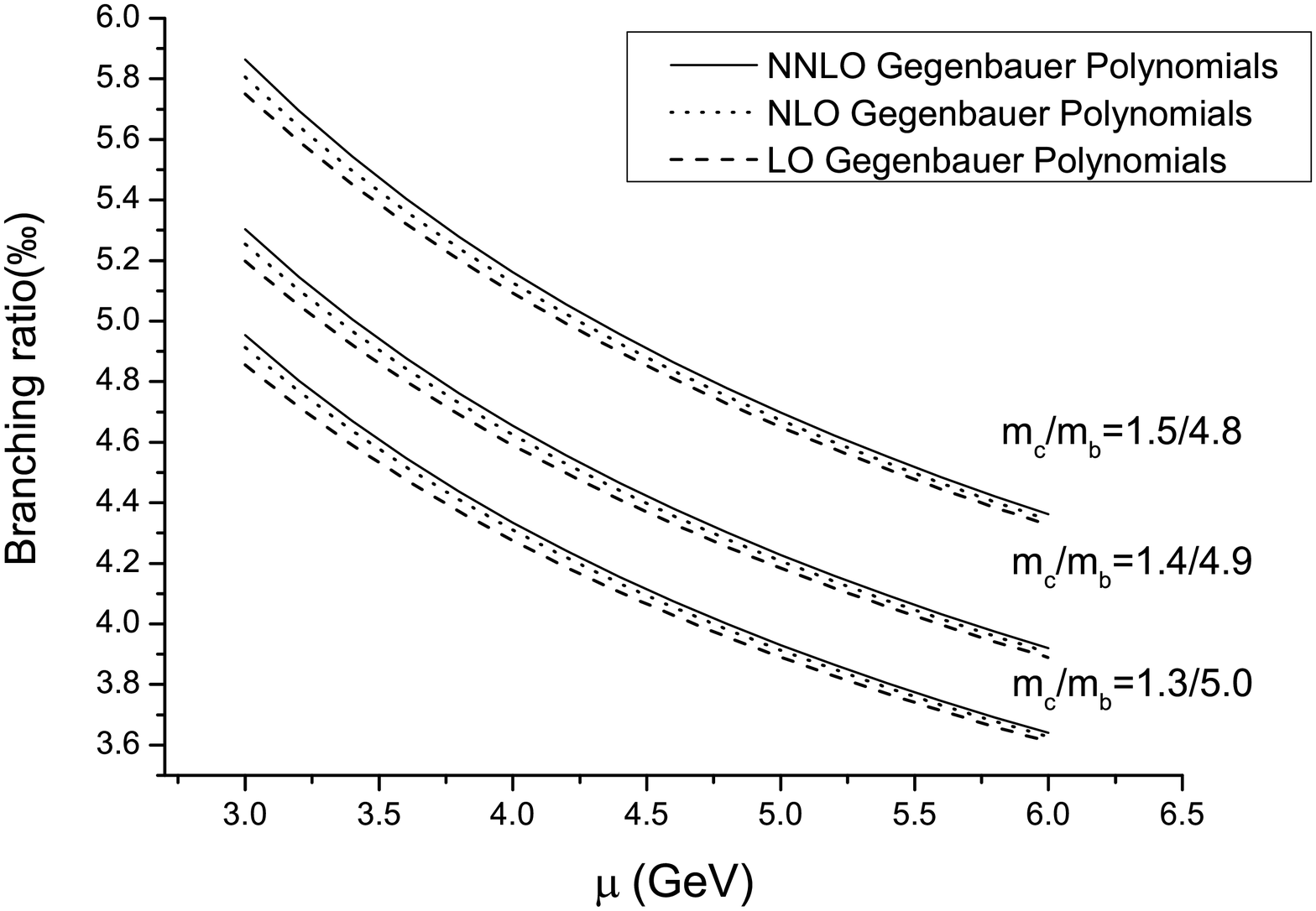}%
\hspace{5mm}
\includegraphics[width=0.480\textwidth]{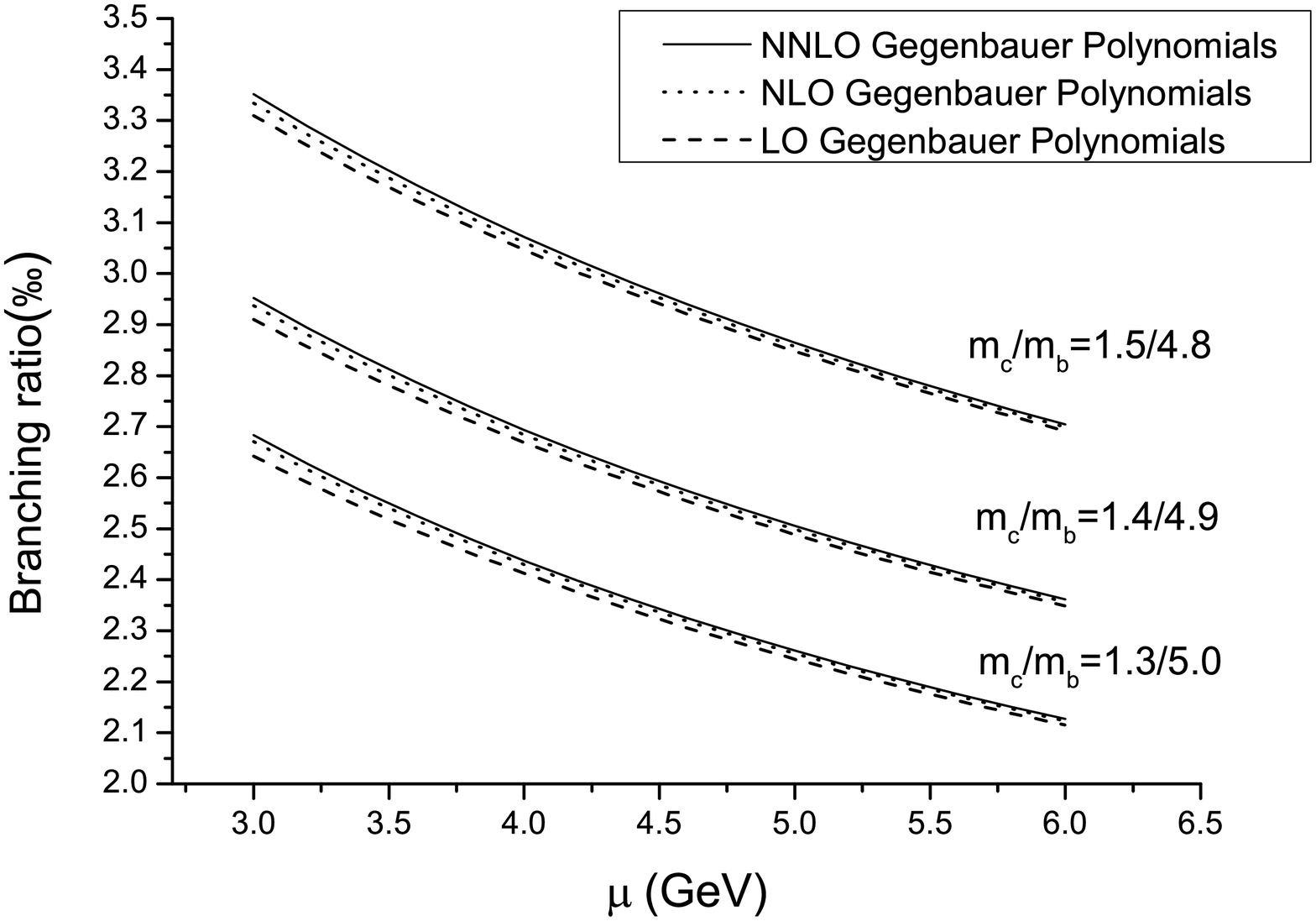}\hspace*{\fill}
\caption{\small The branching ratio of $B_{c}\to \eta_{c} \pi$(left)
and $B_{c}\to J/\psi \pi$(right) versus renormalization scale $\mu$,
for different choices of quark mass. And LO, NLO, NNLO Gegenbauer
polynomials of Pion's light cone distribution amplitude are
considered respectively.} \label{Fig:Gegenbauer} \vspace{-0mm}
\end{figure}

Apart from the uncertainty of renormalization scale, we also study
the uncertainty from quark mass. We found that both of them are
important for the final results. The vivid figures considering
 both dependence are drawn in Figure~\ref{Fig:Gegenbauer}. In which, we
also detailed the influences from Gegenbauer polynomials of light
cone distribution amplitude of Pion, however, which brings about
slight influence to the final result.

After considered the uncertainties stated above,  we give out our
results based on NR QCD factorization in Table~\ref{tab:Bc-nonlep1}
and compare them with that calculated from other models. The LO
results are generally close to results of the QCD sum
rule\cite{KKL,exBc}, the constituent quark
model\cite{Chang:1992pt,AbdEl-Hady:1999xh,Ivanov:2006,Junfeng:2009}
and light-front ISGW model\cite{narod}, however larger than that of
the relativistic potential model\cite{CdF} and the relativistic
quark model \cite{Faust}.  Our work showed that the NLO corrections
substantially enhance the branching ratios, and the NLO QCD
correction K factors are, $1.75^{+0.14+0.18}_{-0.34-0.11}$ for
$\Gamma(B_c\to \eta_{c}\pi)$ and $1.31^{+0.06+0.18}_{-0.18-0.12}$
for $\Gamma(B_c\to J/\psi\pi)$.

Moreover, we want to study the degree of importance  for the
factorizable part at NLO accuracy. After calculation, we found that
the asymptotic factorizable contribution can be well-represented the
majority of the branching ratio. To present it more vividly, let us
set $m_c=1.4\mathrm{GeV},~m_b=4.9\mathrm{GeV}$, and
$\mu=3\mathrm{GeV}$, and we obtain
\begin{eqnarray}
&&Br(B_c\to
\eta_{c}\pi)^{\mathrm{Asymp.~factorizable}}=5.10\permil,~~~~~Br(B_c\to
\eta_{c}\pi)^{\mathrm{total}}=5.20\permil\,,\nonumber\\
&&Br(B_c\to
J/\psi\pi)^{\mathrm{Asymp.~factorizable}}=3.06\permil,~~~Br(B_c\to
J/\psi\pi)^{\mathrm{total}}=2.91\permil\,.\label{eq:fact-count}
\end{eqnarray}

Experimentally, the $pp$ collisions at LHC have been performed at
center-of-mass energy $\sqrt{s}=8\mathrm{TeV}$. And the energy will
arrive at $\sqrt{s}=14\mathrm{TeV}$ in the future days. $pp$
collisions provide a mass-production source for $B_c$ meson. Since
the  $B_c^*$ decays into the  $B_c$ with a probability of almost
100\%, including the contributions from the S-wave excited states,
the cross section of the $B_c$ meson at LHC was estimated to be
around $10^2\mathrm{n b}$. With $10\mathrm{fb}^{-1}$ of integrated
luminosity, there are around $10^9$ events for $B_c$ production.
Then, the measurements of $B_c\to J/\psi\pi\to \mu^+\mu^- \pi$ and
$B_c\to J/\psi\pi\to e^+ e^- \pi$ are feasible, and the events are
presented in Table~\ref{tab:events}, in which we considered the
quark mass dependence.

At last, let us study some certain channels which have been measured
by the LHCb collaboration recently. The LHCb collaboration have
measured $Br(B_{c}^{+}\to J/\psi \pi^+\pi^-\pi^+)/Br(B_{c}^{+}\to
J/\psi \pi^+)$ to be $2.41\pm0.30\pm0.33$, using $0.8fb^{-1}$  data
of pp collisions at center-of-mass energy
$\sqrt{s}=7\mathrm{TeV}$\cite{LHCb:2012}. Experimentally the
reconstructed invariant mass distribution of the $\pi^+\pi^-\pi^+$
combinations favors a resonance state $a_1^+(1260)$.

In theoretical aspects, there are  mainly two channels:
$B_{c}^{+}\to J/\psi a_1^+(1260)$ followed with $a_1^+(1260)\to
\pi^+\pi^-\pi^+$; and $B_{c}^{+}\to \Psi(2S) \pi^+$ with $\Psi(2S)
\to J/\psi \pi^+\pi^-$ which contribute to the signal of
$B_{c}^{+}\to J/\psi \pi^+\pi^-\pi^+$. Practically, $Br(B_{c}^{+}\to
\Psi(2S) \pi^+)/Br(B_{c}^{+}\to J/\psi \pi^+)=
(A^2_0(0)_{\Psi(2S)}(m_{B_c}^2-m_{\psi(2S) }^2){}^3)/$
~$(A^2_0(0)_\Psi(m_{B_c}^2-m_{\psi }^2){}^3)\approx0.26$, and
$Br(\Psi(2S) \to J/\psi \pi^+\pi^-)=33.6\%$\cite{PDG:2010}. So the
contribution to $Br(B_{c}^{+}\to J/\psi
\pi^+\pi^-\pi^+)/Br(B_{c}^{+}\to J/\psi \pi^+)$ from $\Psi(2S)$ is
0.08, which explained the experiment data that favor the resonance
state $a_1^+(1260)$ rather than $\Psi(2S)$.

In the former section, we have performed a complete QCD NLO
calculation of $B_c$ decays into S-wave charmonia and light mesons.
And we found that the factorizable contribution account for more
than 85\% of total results at NLO accuracy in heavy quark limit as
Eq.~(\ref{eq:fact-count}). Here we assume that it is also hold in
$B_{c}^{+}\to J/\psi a_1^+(1260)$, i.e. it can reserve a
high-accuracy just considering the factorizable diagrams. So we
adopt the naive factorization scheme. And for the axial-vector meson
$a_1^+(1260)$, the matrix element for its creation is
\begin{eqnarray}
\langle a_1^+(1260)(p,\epsilon^*)\vert\bar u\gamma_{\mu}\gamma_{5}
d\vert 0\rangle &=&-if_{a_1}m_{a_1}\epsilon^*_{\mu}\,.
\end{eqnarray}

\begin{table}[t]
\caption{\label{tab:Bc-nonlep1}
         Branching ratios (in \permil) of
         exclusive non-leptonic  $B_c$ decays into S-wave charmonium
         states. For the lifetime of the $B_c$ we take $\tau(B_c) =0.453$ ps. In
         our work, we choose the quantities
$m_c=1.4~\mathrm{GeV}$, $m_b=4.9~\mathrm{GeV}$, and
$\mu=3~\mathrm{GeV}$. The uncertainty in the first column of the
value is from varying  the renormalization scale $\mu$ from
$2.5~\mathrm{GeV}$ to $5~\mathrm{GeV}$; while the uncertainty in the
second column comes from varying the quark mass $m_c/m_b$ from
$1.5/4.8$ to $1.3/5.0$.}
\begin{center}
\begin{tabular}{|c||c|c|c|c|c|c|c|c|c|c|}
\hline
 Mode & This work(NLO) & LO & \cite{KKL,exBc} & \cite{Chang:1992pt} &
 \cite{narod} & \cite{CdF} & \cite{Faust} & \cite{AbdEl-Hady:1999xh} & \cite{Ivanov:2006} & \cite{Junfeng:2009}\\
\hline
 $B_c^+ \to \eta_c \pi^+$  & $5.19^{+0.44+0.55}_{-1.01-0.34}$  & 2.95 & 2.0  & 2.2  & 1.3  &
0.26 & 0.85 & 1.4   & 1.9 & 2.1\\
\hline $B_c^+ \to \eta_c \rho^+$  & $14.5^{+1.29+1.53}_{-2.92-0.95}$
&7.89 & 4.2 & 5.9  &  3.0 & 0.67
 & 2.0 &  3.3  & 4.5 & -\\
\hline $B_c^+ \to \eta_c K^+$  & $0.38^{+0.03+0.04}_{-0.07-0.02}$ &
0.21 & 0.13&0.17 & 0.13  &0.02
 &  0.06&  0.11  & 0.15 &- \\
\hline $B_c^+ \to \eta_c K^{*+}$  & $0.77^{+0.07+0.08}_{-0.16-0.05}$
&0.41 & 0.20  &  0.31 & 0.21  &0.04
 & 0.11 & 0.18   & 0.25 &- \\
\hline
 $B_c^+ \to J/\psi \pi^+$   &  $2.91^{+0.15+0.40}_{-0.42-0.27}$ &2.22 & 1.3  & 2.1  & 0.73 &
1.3 & 0.61 & 1.1& 1.7 &2.0 \\
\hline $B_c^+ \to J/\psi \rho^+$  &$8.08^{+0.45+1.09}_{-1.21-0.73}$
& 6.03& 4.0  & 6.5  & 2.1 &
3.7 & 1.6 & 3.1& 4.9 &- \\
\hline $B_c^+ \to J/\psi K^+$  &$0.22^{+0.01+0.03}_{-0.03-0.02}$
&0.16 & 0.11  & 0.16  & 0.07 &
0.07 & 0.05 & 0.08& 0.13 & - \\
\hline $B_c^+ \to J/\psi K^{*+}$  &$0.43^{+0.02+0.06}_{-0.07-0.04}$
& 0.32& 0.22  & 0.35  & 0.16 &
0.20& 0.10 & 0.18& 0.28 & - \\
\hline
 $B_c^+ \to \psi(2S) \pi^+$   &  $0.76^{+0.04+0.10}_{-0.11-0.07}$ &0.58 & -  & 0.27  & - &
0.19 & 0.11 & -& - & - \\
\hline $B_c^+ \to  \psi(2S) \rho^+$
&$2.11^{+0.12+0.28}_{-0.32-0.19}$ & 1.57& -  & 0.77  & - &
0.48 & 0.18 & -& - &- \\
\hline $B_c^+ \to  \psi(2S) K^+$
&$0.057^{+0.003+0.008}_{-0.008-0.005}$ &0.042 & -  & 0.019  & - &
0.009 & 0.01 & -& - & - \\
\hline $B_c^+ \to  \psi(2S) K^{*+}$
&$0.112^{+0.005+0.015}_{-0.018-0.010}$ & 0.083& -  & 0.041  & -&
0.026& 0.01 & -& - & - \\
\hline
\end{tabular}
\end{center}
\end{table}
\begin{table}
\begin{center}
\caption{\label{tab:events}The events  of $B_c\to J/\psi\pi\to
\mu^+\mu^- \pi$, $B_c\to J/\psi\pi\to e^+ e^- \pi$ and $B_c\to
\eta_c\pi\to \gamma\gamma \pi$ with $10\mathrm{fb}^{-1}$ data, using
various values of the $c$-quark mass $m_c$ and fixed $b$-quark mass
$m_b=4.8$ GeV.}
\begin{tabular}{|c||c|c|c|c|c||c|c|c|c|c|}
\hline - & \multicolumn{5}{|c||}{~~~Tevatron~($\sqrt S=2.$
TeV)~~~}& \multicolumn{5}{|c|}{~~~LHC~($\sqrt S=14.$ TeV)~~~}\\
\hline $m_c$~(GeV) & $~1.4~$& $~1.5~$ & $~1.6~$ &~$1.7$~&~$1.8$~&
~$1.4$~ & ~$1.5$~ & ~$1.6$~
&~$1.7$~&~$1.8$~\\
\hline $\sigma_{B_{c}}~(nb)$ & 13.4 & 10.5 & 8.48 & 6.89 & 5.63&
214 & 160 & 139 & 114& 95.1\\
\hline $\mu^+\mu^- \pi~(\times10^4)$ & 2.54 & 2.15 &1.79  & 1.56
&1.40 &
 40.6& 32.8 & 29.3& 25.9 &23.6  \\
 \hline $e^+ e^- \pi~(\times10^4)$ & 2.54 & 2.15 &1.79  & 1.57 &1.40 &
 40.6& 32.8 & 29.3& 26.0 &23.7  \\
 \hline $\gamma\gamma \pi$ & 47 & 37 &32  &  27&24 &
 754& 567 & 525& 459 & 413 \\
\hline
\end{tabular}
\end{center}
\end{table}
Then we can obtain the ratio
\begin{eqnarray}
\frac{ Br(B_{c}^{+}\to J/\psi a_1^+(1260))}{Br(B_{c}^{+}\to J/\psi
\pi^+)} &=&\frac{f^2_{a_1}\lambda_0}{f^2_\pi A^2_0(0)}[\lambda_1
A^2_1(m^2_{a_1})+\lambda_2 A^2_2(m^2_{a_1})+\lambda_3
A_1(m^2_{a_1})A_2(m^2_{a_1})-\lambda_4 V^2(m^2_{a_1})]\,,\nonumber
\end{eqnarray}
with
\begin{eqnarray}
\lambda_0 &=&\frac{\sqrt{\left(m_{a_1}^2+m_{B_c}^2-m_{\psi
}^2\right){}^2-4 m_{a_1}^2 m_{B_c}^2}}{4 m_{\psi }^2
   \left(m_{B_c}+m_{\psi }\right){}^2 \left(m_{B_c}^2-m_{\psi }^2\right){}^3}\,,
\end{eqnarray}
\begin{eqnarray}
\lambda_1 &=&\left(m_{B_c}+m_{\psi }\right){}^4 \left(-2 m_{a_1}^2
\left(m_{B_c}^2-5 m_{\psi
   }^2\right)+m_{a_1}^4+\left(m_{B_c}^2-m_{\psi }^2\right){}^2\right)\,,
\end{eqnarray}
\begin{eqnarray}
\lambda_2 &=&\left(-2 m_{a_1}^2 \left(m_{B_c}^2+m_{\psi
}^2\right)+m_{a_1}^4+\left(m_{B_c}^2-m_{\psi
}^2\right){}^2\right){}^2\,,
\end{eqnarray}
\begin{eqnarray}
\lambda_3 &=&2 \left(m_{B_c}+m_{\psi }\right){}^2
\left(m_{a_1}^2-m_{B_c}^2+m_{\psi }^2\right)
   \left(m_{a_1}^2-\left(m_{B_c}-m_{\psi }\right){}^2\right) \left(m_{a_1}^2-\left(m_{B_c}+m_{\psi }\right){}^2\right)\,,
\end{eqnarray}
\begin{eqnarray}
\lambda_4 &=&-8 m_{a_1}^2 m_{\psi }^2 \left(-2 m_{a_1}^2
\left(m_{B_c}^2+m_{\psi }^2\right)+m_{a_1}^4+\left(m_{B_c}^2-m_{\psi
   }^2\right){}^2\right)\,.
\end{eqnarray}

According to the Ref.~\cite{Barbar:2007}, we assume
$Br(a_1^+(1260)\to \pi^+\pi^-\pi^+)$ is equal $Br(a_1^+(1260)\to
\pi^+\pi^0\pi^0)$ and its value is 50\%.  And we take the input
parameters $f_{a_1}=0.23\mathrm{GeV}$ from the QCD sum
rules\cite{fa1:2007}, $m_{a_1}=1.23\mathrm{GeV}$ from the Particle
Data Group\cite{PDG:2010}. The final result is
\begin{eqnarray}
\frac{ Br(B_{c}^{+}\to J/\psi \pi^+\pi^-\pi^+)}{Br(B_{c}^{+}\to
J/\psi \pi^+)} &=&2.75^{+0.03+0.22}_{-0.04-0.05}\,,\nonumber
\end{eqnarray}
The uncertainties of our result come from the renormalization scale
and quark mass. It is compatible with the experimental data
$2.41\pm0.30\pm0.33$  when considering its uncertainty.

In order to conveniently compare  with the LHCb's data, we present
our prediction and experimental data in Table~\ref{tab:comparision}.
For the former three channel, our results can explain the data
perfectly. While for the latter three channels, more data is needed
to investigate the validity of NRQCD factorization on $B_c$ decays .

\begin{table}
\centering
 \caption{\label{tab:comparision}Branching fraction ratio in comparison with the LHCb's data.}
\hspace{-1.5mm}
\begin{tabular}{|c||c|c|}\hline

Ratio & our work & LHCb
\\\hline $ Br(B_{c}^{+}\to J/\psi \pi^+\pi^-\pi^+)/Br(B_{c}^{+}\to
J/\psi \pi^+)$  & 2.75 & $2.41\pm0.30\pm0.33$\cite{LHCb:2012}
\\\hline$Br(B_{c}^{+}\to J/\psi K^+)/Br(B_{c}^{+}\to
J/\psi \pi^+)$  & 0.075 & $0.069\pm0.019\pm0.005$\cite{Aaij:2013vcx}
\\\hline
$ Br(B_{c}^{+}\to \Psi(2S)\pi^+)/Br(B_{c}^{+}\to J/\psi \pi^+)$ &
0.260 & $0.250\pm0.068\pm0.014\pm0.006$\cite{LHCb:2013}
\\\hline $ Br(B_{c}^{+}\to \Psi(2S)K^+)/Br(B_{c}^{+}\to J/\psi \pi^+)$ &
0.02 & -
\\\hline$ Br(B_{c}^{+}\to J/\psi \rho^{+})/Br(B_{c}^{+}\to J/\psi
\pi^+)$  & 2.77&-
\\\hline
$ Br(B_{c}^{+}\to J/\psi K^{*+})/Br(B_{c}^{+}\to J/\psi \pi^+)$  &
0.147 & -
\\\hline
\end{tabular}
\end{table}

\section{Conclusions\label{sect:con}}
We have performed a comprehensive NLO analysis for the $B_c$ meson
decays into S-wave charmonia and light mesons such as $\pi$, $\rho$,
$K$ and $K^*$. The NLO QCD correction provides a large K factor
which substantially enhance the branching ratio, while  the $\mu$
dependence is reduced corresponding. Considering about uncertainties
of sorts of input parameters, we find out the largest uncertainty
comes from the masses of bottom and charm quarks.

In the heavy quark limit,  the analytic amplitude up to NLO accuracy
is derived. Therein logarithm $\ln z$ with $z=m_c/m_b$ is absent in
the contribution for color-singlet operator, while this kind of
logarithm and double logarithm $\ln^2 z$ emerge in that of
color-octet operator. And the result of asymptotic NLO where we only
reserve the leading order in the $z$ expansion can account for more
than $85\%$ of the complete NLO's result in which $z$ is fixed to
its physical value. Therefor it is enough to use the asymptotic
formulas for phenomenological studies at NLO accuracy.

Numerical results show that the latest LHCb's data on $B_c$ decays
can be explained perfect using NRQCD factorization under their
corresponding uncertainties. We also predicted another three
channels which shall be checked in the upcoming data. The large
branching ratio and the clear signal of final states make it
reliable for the measurement of the absolute branching ratios for
the processes $B_{c}\to J/\psi \pi$, $B_{c}\to J/\psi \rho$ and
$B_{c}\to J/\psi K$ within the updated LHCb's data.

\section*{Acknowledgement}
\hspace{2cm}

This work was supported in part by the National Natural Science
Foundation of China(NSFC) under the grants 10935012, 10821063 and
11175249.
\section*{Appendix}

\begin{appendix}
\section{LCDA for light mesons and
projection operators for heavy quarkonia \label{appendix A}}

Considering the twist-2 and twist-3 light-cone distribution
amplitudes for  Pion, we have the matrix element of quarks
hadronization projection operator\cite{Beneke:2000,Beneke:2001}
\footnotesize\begin{equation} \label{curproj} \bar{u}_{\alpha
a}(xP)\Gamma(x,\ldots)_{\alpha\beta,ab,\ldots} v_{\beta b}(\bar{x}P)
\,\longrightarrow\, \frac{i f_\pi}{4 N_c}\int_0^1
dx\,\,M^\pi(x)_{\alpha\beta}\Gamma(x,\ldots)_{\alpha\beta,aa,\ldots}\,,
\end{equation}\normalsize
with the decay constant $f_\pi=130.4\mathrm{MeV}$, $\bar{x}=1-x$ and

\begin{equation}
M^\pi(x)_{\alpha\beta} =
 \Bigg\{ \not\!P \gamma_5 \, \phi(x)  -
   \mu_\pi \gamma_5 \left(  \phi_p(x)
       - i \sigma_{\mu\nu}\, n_-^\mu v^\nu \,
       \frac{\phi^\prime_\sigma(x)}{6}
 +i \sigma_{\mu\nu}  P{}^\mu \, \frac{\phi_\sigma(x)}{6}
       \, \frac{\partial}{\partial
         k_\perp{}_{\nu}} \right)
 \Bigg\}_{\alpha\beta},
\label{pimeson2}
\end{equation}
here $\mu_\pi=m_\pi^2/(m_u+m_d)$, $n_{\pm}$ are light cone vectors,
$\phi(x)$  is twist-2 distribution amplitude and  $\phi_p(x)$ and
$\phi_\sigma(x)$ are twist-3 ones. For Pion, up to twist-2
\cite{LCDA, Beneke:2000} \footnotesize
\begin{eqnarray}
\phi_\pi(x)=6x\bar{x}\{1+a_1C^{3/2}_2(\bar{x}-x)+a_2C^{3/2}_4(\bar{x}-x)\},
\end{eqnarray}\normalsize
with $a_1=0.44$, $a_2=0.25$, and the Gegenbauer polynomials are
defined by \footnotesize
\begin{eqnarray}
C^{3/2}_2(z)=\frac{3}{2}(5z^2-1),~~~~C^{3/2}_4(z)=\frac{15}{8}(21z^4-14z^2+1)\,.
\end{eqnarray}\normalsize

Then, for  vector meson $\rho$, the corresponding matrix element of
hadronization projection operator is\cite{Beneke:2000,Beneke:2001}
\footnotesize\begin{equation} \label{curproj} \bar{u}_{\alpha
a}(xP')\Gamma(x,\ldots)_{\alpha\beta,ab,\ldots} v_{\beta
b}(\bar{x}P') \,\longrightarrow\, \frac{i f_\rho}{4 N_c}\int_0^1
dx\,\,M^\rho(x)_{\alpha\beta}\Gamma(x,\ldots)_{\alpha\beta,aa,\ldots}\,,
\end{equation}\normalsize
\begin{equation}
  M_{\alpha\beta}^\rho =  M_{\alpha\beta}^{\rho}{}_\parallel +
                          M_{\alpha\beta}^{\rho}{}_\perp\,,
\label{rhomeson2}
\end{equation}
with
\begin{eqnarray} M^{\rho}_\parallel &=& -\frac{if_\rho}{4} \,
\frac{m_\rho(\varepsilon^*\cdot
  n_+)}{2 E} \,E\,
 \slash n_- \,\phi_\parallel(u)
 -\frac{if_\perp m_\rho}{4}  \,\frac{m_\rho(\varepsilon^*\cdot n_+)}{2 E}
 \, \Bigg\{-\frac{i}{2}\,\sigma_{\mu\nu} \,  n_-^\mu  n_+^\nu \,
 h_\parallel^{(t)}(u)
\nonumber\\[0.1cm]
&& \hspace*{-0.0cm} - \,i E\, \int_0^u dv \,(\phi_\perp(v) -
h_\parallel^{(t)}(v)) \
     \sigma_{\mu\nu}   n_-^\mu
     \, \frac{\partial}{\partial k_\perp{}_\nu}
  +\frac{h_\parallel'{}^{(s)}(u)}{2}\Bigg\}\, \Bigg|_{k=u p'}\,,
\end{eqnarray}
and
\begin{eqnarray}
   M^{\rho}_\perp &=& -\frac{if_\perp}{4} \,
   E\,\slash \varepsilon^*_\perp \slash n_- \, \phi_\perp(u)
 -\frac{if_\rho m_\rho}{4} \,\Bigg\{\slash \varepsilon^*_\perp\, g_\perp^{(v)}(u)
\nonumber\\[0.1cm] &&
-   \,E \, \int_0^u dv\, (\phi_\parallel(v) - g_\perp^{(v)}(v))  \
       \slash n_- \, \varepsilon^*_{\perp\mu} \,\frac{\partial}{\partial
         k_{\perp\mu}}
\cr && + \,i \epsilon_{\mu\nu\rho\sigma} \,
        \varepsilon_\perp^{*\nu} n_-^\rho\, \gamma^\mu\gamma_5
         \left[ n_+^\sigma \,\frac{g_\perp'{}^{(a)}(u)}{8}-
          E\,\frac{g_\perp^{(a)}(u)}{4} \, \frac{\partial}{\partial
         k_\perp{}_\sigma}\right]
 \Bigg\}
 \, \Bigg|_{k=up'}.
\end{eqnarray}
Up to twist-2, the LCDA for longitudinally polarized $\rho$ meson is
\footnotesize\begin{equation} \label{rhow}
\phi_{\rho,\parallel}(x)=6x\bar{x}\{1+a^\rho_1
C^{3/2}_2(\bar{x}-x)\},
\end{equation}\normalsize
here $a^\rho_1=0.18$.

In addition the twist-2 LCDA for  $K$ meson is\cite{LCDA}
\footnotesize\begin{equation} \label{rhow}
\phi_K(x)=6x\bar{x}\{1+0.51(\bar{x}-x)+0.2C^{3/2}_2(\bar{x}-x)\},
\end{equation}\normalsize
and  for longitudinally polarized $K^*$ meson
\footnotesize\begin{equation} \label{rhow}
\phi_{K^*,\parallel}(x)=6x\bar{x}\{1+0.57(\bar{x}-x)+0.07C^{3/2}_2(\bar{x}-x)\},
\end{equation}\normalsize

 At last, using  leading Fock states for heavy quarkonium, the
quarks hadronizaiton projection operators are\cite{Qiao:2011}
\footnotesize\begin{eqnarray} v(p_b)\,\overline{u}(p_c)&
\longrightarrow& {1\over 2 \sqrt{2}}
\gamma_5(\not\!P_{B_c}+m_b+m_c)\, \times \left( {1\over
\sqrt{\frac{m_b+m_c}{2}}} \psi_{B_c}(0)\right) \otimes \left( {{\bf
1}_c\over \sqrt{N_c}}\right)\label{eq:2},\nonumber\\
v(p_{\bar{c}})\,\overline{u}(p_c)& \longrightarrow& {1\over 2
\sqrt{2}} \not\!\epsilon(\not\!P_{\Psi}+m_c+m_c)\, \times \left(
{1\over \sqrt{\frac{m_c+m_c}{2}}} \psi_{\Psi}(0)\right) \otimes
\left( {{\bf 1}_c\over \sqrt{N_c}}\right).
\end{eqnarray}\normalsize
The NLO Schr\"odinger wave function at origin of $J/\psi$ is
determined by leptonic decay width \footnotesize\begin{eqnarray}
|\psi _{J/\psi}(0)|^2=\frac{m^2_{J/\psi
}}{16\pi\alpha^2e_{c}^2}\frac{ \Gamma (J/\psi\rightarrow e^+
e^-)}{(1+\pi\alpha_s C_F/v-4\alpha_s C_F/\pi)}.
\end{eqnarray}\normalsize

\section{Renormalization  and infre-red subtractions \label{renormal}}
 The renormalization constants include $Z_{2}$,
$Z_{3}$, $Z_{m}$, and $Z_{g}$, corresponding to heavy quark field,
gluon field, quark mass, and strong coupling constant g,
respectively. Here, in our calculation the $Z_{g}$ is defined in the
modified-minimal-subtraction ($\mathrm{\overline{MS}}$) scheme,
while for the other three the on-shell ($\mathrm{OS}$) scheme is
adopted, which tells \footnotesize\begin{eqnarray}
&&\hspace{-0.3cm}\delta Z_m^{OS}=-3C_F
\frac{\alpha_s}{4\pi}\left[\frac{1}{\epsilon_{UV}}-\gamma_{E}+
\ln\frac{4\pi\mu^{2}}{m^{2}}+\frac{4}{3}+{\mathcal{O}}(\epsilon)\right]\;
,\nonumber
\\ &&\hspace{-0.3cm}\delta Z_2^{OS}=-C_F
\frac{\alpha_s}{4\pi}\left[\frac{1}
{\epsilon_{UV}}+\frac{2}{\epsilon_{IR}}-3\gamma_{E}+3\ln
\frac{4\pi\mu^{2}}{m^{2}}+4+{\mathcal{O}}(\epsilon)\right]\; ,\nonumber\\
&&\hspace{-0.3cm}\delta Z_3^{OS}= \frac{\alpha_s}{4\pi}
\left[(\beta_0-2C_A)(\frac{1}{\epsilon_{UV}}-
\frac{1}{\epsilon_{IR}})+{\mathcal{O}}(\epsilon)\right]\; ,\nonumber\\
&&\hspace{-0.3cm}\delta Z_g^{\overline{MS}}=-\frac{\beta_0}{2}
\frac{\alpha_s}{4\pi}\left[\frac{1}{\epsilon_{UV}}-\gamma_{E}+
\ln4\pi+{\mathcal{O}}(\epsilon)\right]\; .\label{eq:13}
\end{eqnarray}\normalsize
While for light quark such as u and d quarks, the corresponding
renormalization constant is \footnotesize\begin{equation}\label{}
    \delta Z_2^{OS}=-C_F \frac{\alpha_s}{4\pi}(\frac{1}{\epsilon_{UV}}
    -\frac{1}{\epsilon_{IR}})\,.
\end{equation}\normalsize
On above, $\delta Z_{i}=Z_{i}-1$, and
$\beta_{0}=(11/3)C_{A}-(4/3)T_{f}n_{f}$ is the one-loop coefficient
of the QCD beta function; $C_{A}=3$ and $T_{F}=1/2$ attribute to the
SU(3) group; $\mu$ is the renormalization scale.

We write the renormalized operator matrix elements
as\cite{Bell:2007} \footnotesize\begin{align} \langle Q_{i}
\rangle_\text{ren} &= Z_\psi \, \hat{Z}_{i j} \, \langle Q_{j}
\rangle_\text{bare}\,, \label{eq:defQren}
\end{align}\normalsize
where $i, j=0, 8$ and $Z_\psi = Z_b^{1/2}Z_c^{1/2} Z_q$ contains the
quark field renormalization factors of the massive b-quark $Z_b$,
the massive c-quark $Z_c$ and the massless quarks $Z_q$, whereas
$\hat{Z}$ is the operator renormalization matrix in the effective
theory. It reads \footnotesize\begin{align} \hat{Z} &= 1+\left(
\begin{array}{c c}
\rule[-2mm]{0mm}{7mm} 0 & 6  \\
\rule[-2mm]{0mm}{7mm} \frac{4}{3}& -2
\end{array}
\right)\frac{\alpha_s}{4\pi}\frac{1}{\epsilon}\,.\label{eq:Z1}
\end{align}\normalsize

All of  soft IR divergences are canceled when summing them up, and
Coulomb divergences can be canceled by the corresponding
counter-term from the NLO Schr\"odinger wave function at origin.
While the left collinear divergences can be removed by Pion wave
function's subtraction\cite{Braaten}.

\section{Formulas for non-factorizable contribution}
In this subsection, the asymptotic formulas for one loop
non-factorizable  contribution are presented, where $x$ is the
collinear quark's momentum fraction in Pion and $z=m_c/m_b$ is the
mass ratio for charm quark and bottom quark. The results are valid
in heavy quark limit.
 \footnotesize
\begin{eqnarray}
T_{\mathrm{nf},0,x}^{(1)}(\eta_c)&=&\phi_\pi(x)\{\frac{2} { x
}\ln(\frac{m_b^2}{\mu^2})+\frac{2 (x-1) \ln ^2(x)}{3 x (2
x-1)}-\frac{2
   (\text{Li}_2(\frac{x-1}{x})-\text{Li}_2(\frac{x}{x-1}))}{3 x}\nonumber\\
&&-\frac{1}{3 (x-1) x (2 x-1)}f_1-\frac{1}{3 x (2 x-1)}f_2+\frac{1}{3 (2 x-1)^3}f_3 +\frac{1}{3 (x-1) x (2 x-1)^3}f_4\nonumber\\
&&+\frac{1}{3 x^2 (2 x-1)^3}f_5+\frac{1}{3 (x-1) x^2 (2
x-1)^3}f_6\}\,,
\end{eqnarray}
\begin{eqnarray}
T_{\mathrm{nf},0,x}^{(1)}(\Psi)&=&T_{\mathrm{nf},0,x}^{(1)}(\eta_c)\,,
\end{eqnarray}
with
\begin{eqnarray}
f_1&=& 2  (x^2 (\ln (2)-4)-x (2+2 \ln (2))+2+\ln (2) ) \ln
(x+1)\,,\nonumber
\end{eqnarray}
\begin{eqnarray}
f_2 &=& 2 (x-1)  (-\text{Li}_2 (\frac{x+1}{2 x^2} )+\text{Li}_2
(-\frac{1}{2 x} )+2
   \text{Li}_2 (\frac{1}{x} )-\text{Li}_2(-2 x)+\text{Li}_2(2 (x+1)) )\,,\nonumber
\end{eqnarray}
\begin{eqnarray}
f_3 &=& 8 (x-1)^2  (\text{Li}_2(4-2 x)+2 \text{Li}_2 (\frac{1}{1-x}
)-\text{Li}_2 (-\frac{x-2}{2
   (x-1)^2} )+\text{Li}_2 (\frac{1}{2 (x-1)} )-\text{Li}_2(2 x-2) )\,,\nonumber
\end{eqnarray}
\begin{eqnarray}
f_4 &=& \ln (x) (-4  (2 x^2-3 x+1 )^2 \ln (x+1)-8 x
(x-1)^3 \ln (2-x)+3+6 \ln (2)\nonumber\\
&& +x (2 x (2 x (-2 x+2 (5 x-14) \ln
   (2)-3)+15+56 \ln (2))-17-46 \ln (2)))\,,\nonumber
\end{eqnarray}
\begin{eqnarray}
f_5&=& \ln (1-x) (x (x (4 x (x-10 x \ln (2)+5+18 \ln (2))-37-38
\ln (2))+19+4 \ln (2))\nonumber\\
&&+2 (x-1) x (\ln (x+1)+4 (x-1)
   x (2 \ln (2-x)+\ln (x+1)))-3)\,,\nonumber
\end{eqnarray}
\begin{eqnarray}
f_6&=& -8 x^2 (x-1)^3 \ln ^2(1-x)+(x-1)  (x  (x  (x^2 (20+8 \ln
(2))-x (72+16 \ln (2))\nonumber\\
&&+79+8 \ln
   (2) )-32 )+4 ) \ln (2-x)+x  (2 x  (2 x  (28 x^2-58 x+45 )-31 )\nonumber\\
&&+\pi ^2 (x-1)^2+x
   (103-2 x (6 x (6 x-13)+83)) \ln (2)+8-33 \ln (2) )+4 \ln (2)\,.\nonumber
\end{eqnarray}
\begin{eqnarray}
T_{8,nf,x}^{(1)}(\eta_c)&=&\phi_\pi(x)\{\frac{-11 C_A+48 x N_c+2
n_f-6}{9 x}\ln(\frac{m_b^2}{\mu^2})
+\frac{9 (x-1) \ln (z)-5 x+2}{3 (x-1) x}C_F-\frac{ (3 C_A-1 ) \ln ^2(z)}{18 x}\nonumber\\
&&-\frac{2 (-3 \ln (z)+5+3 \ln (2))}{27 x}n_f-\frac{\ln (z) }{54
(x-1) x}((138 x -144 x \ln (2) -138 +90 \ln (2)) C_A+144 x\nonumber\\
&&+96 x^2
   \ln (1-x)-96 x^2 \ln (x)-96 x \ln (1-x)+96 x \ln (x)+372 x \ln
   (2)-144-210 \ln (2))\nonumber\\
&&+\frac{1}{N_c}(\frac{2
   (\text{Li}_2(\frac{x-1}{x})-\text{Li}_2(\frac{x}{x-1}))}{3 x}-\frac{(x-2) (x (10 x-11)+2) \ln  (1-\frac{x}{2} )}{3 (1-2 x)^2 x^2 }\nonumber\\
&&+\frac{1}{3 x (2 x-1)}f_2-\frac{1}{3 (2 x-1)^3}f_3+\frac{1}{3
(x-1) x (2
   x-1)^3 }f_7+\frac{1}{3 x^2 (2
   x-1)^3 }f_8+\frac{1}{3 (x-1) x (2 x-1)^3 }f_9)\nonumber\\
&&+C_A(-\frac{(2 x+1)
    (\text{Li}_2 (\frac{x-1}{x} )-\text{Li}_2 (\frac{x}{x-1}
   ) )}{3 x}-\frac{  (x^2 \text{Li}_2 (\frac{1-x}{2} )-x^2
   \text{Li}_2 (\frac{x}{2} )+2 x
   \text{Li}_2 (\frac{x}{2} )-\text{Li}_2 (\frac{1-x}{2} ) )
   }{(x-1) x}+\frac{1}{6x}f_2\nonumber\\
&&-\frac{x-1}{3 (2 x-1)^3}f_3+\frac{ (-4 x^2+x-1 )\ln (x) \ln (2
x+1)}{3 x (2 x-1)} -\frac{(x-2) (3 x-2)  (8 x^2-6 x+3 ) C_A \ln
    (1-\frac{x}{2} )}{6 (1-2 x)^2 (x-1) x}\nonumber\\
&&+\frac{ \ln (x) }{6 (x-1) x (2 x-1)^3}f_{10}+\frac{\ln (1-x) }{6
x^2 (2 x-1)^3}f_{11}
-\frac{1}{6 (x-1) x^2 (2 x-1)^3}f_{12}+\frac{1}{108 (x-1) x^2 (2 x-1)^3}f_{13})\nonumber\\
&&-\frac{ (x^2-x+2 )
    (\text{Li}_2(1-x)-\text{Li}_2 (1-\frac{x}{2} ) )}{9 (x-1) x}
   +\frac{(x-3)  (\text{Li}_2(x)-\text{Li}_2 (\frac{x+1}{2} ) )}{9 x}-\frac{3 (x-2)
   \text{Li}_2 (\frac{x}{2} )}{x-1}
   \nonumber\\
&&-\frac{7}{3}
    (\text{Li}_2 (\frac{x-1}{x} )-\text{Li}_2 (\frac{x}{x-1}
   ) )+\frac{3 (x+1) \text{Li}_2 (\frac{1-x}{2} )}{x}+\frac{(x-2)  (78 x^2-51 x-2 ) \ln  (1-\frac{x}{2} )}{18 (x-1)
   x^2}\nonumber\\
&&+\frac{\ln (x) }{18 (x-1) x}f_{14}+\frac{\ln (1-x) }{9 (x-1)
x^2}f_{15}+\frac{6 }{108 (x-1) x}f_{16}\}\,,
\end{eqnarray}

\begin{eqnarray}
T_{8,nf,x}^{(1)}(\Psi)&=&T_{8,nf,x}^{(1)}(\eta_c)+\phi_\pi(x)\{-\frac{16}{3}
 \ln(\frac{m_b^2}{\mu^2})+\ln(z)(\frac{16 (x-1) x
\ln (1-x)-16 (x-1) x \ln (x)+36 x \ln (2)-18 \ln (2)}{9 (x-1)
x}\nonumber\\
&&-\frac{2 (2 x-1) \ln (2) C_A}{3 (x-1) x})+\frac{C_A }{18 (x-1)
x}f_{17}+\frac{1}{18 (x-1) x}f_{18}\}\,,
\end{eqnarray}
with
\begin{eqnarray}
f_7&=& \ln (x)  (4  (2 x^2-3 x+1 )^2 \ln (x+1)+(x+3) (2 x-1)^3+8 x
   (x-1)^3 \ln (2-x)\nonumber\\
&&-2 (4 x (x (5 x-9)+5)-3) (x-1) \ln (2) )\,,\nonumber
\end{eqnarray}
\begin{eqnarray}
f_8&=& \ln (1-x) (x (x (-4 x (x+5)+8 x (5 x-9) \ln (2)+37+38 \ln
(2))-19-4 \ln
   (2))\nonumber\\
&&+2 (x-1) x (-\ln (x+1)-4 (x-1) x (2 \ln (2-x)+\ln
(x+1)))+3)\,,\nonumber
\end{eqnarray}
\begin{eqnarray}
f_9&=&24 x^3 \ln (2) \ln (4-2 x)-8 x  (x^3+3 x-1 ) \ln (2) \ln
(2-x)+x
   (4 x^3 (13 \ln (2)-28)-8 x^2 (-29+3 \ln ^2(2)+8 \ln (2))\nonumber\\
&&+15 x
   (\ln (2)-12)+62+8 \ln (2))-\pi ^2 (x-1)^2+8 x (x-1)^3 \ln ^2(1-x)+2 x
   (x (-4 (x-3) x-13)+6) \ln ^2(x)\nonumber\\
&&-2 \ln ^2(x)+4 i \pi  (x-1)^2 \ln (2)+(2 x (x
   (x (x (4\ln (2)-16)+8-12 \ln (2))+12+13 \ln (2))\nonumber\\
&&-2 (5+3 \ln (2)))+2 \ln
   (2)) \ln (x+1)+4 \ln (x+1)-8-3 \ln (2)\,,\nonumber
\end{eqnarray}
\begin{eqnarray}
f_{10}&=&(16 x (x-1)^4 \ln (2-x)+4 (2 x-1)^3 (x-1)^2 \ln (x+1)-2 x
   (x-1) (2 x (2 x (4 x-8-7 \ln (2))+10+17 \ln (2))\nonumber\\
&&-3 (2+5 \ln (2)))+2 (1-2 x)^2
   (x (4 x-1)+1) (x-1) \ln (2 x+1)+1+2 \ln (2))\,,\nonumber
\end{eqnarray}
\begin{eqnarray}
f_{11}&=& (-32 x^2 (x-1)^3 \ln (2-x)-2 x (2 x-1)^3 (x-1) \ln
   (x+1)+x (x (4 x (x (20 x-37-20 \ln (2))\nonumber\\
&&+29+30 \ln (2))-51-52 \ln (2))+12+\ln
   (16))-1)\,,\nonumber
\end{eqnarray}
\begin{eqnarray}
f_{12}&=& (-(x-2) x (2 x-1) (3 x-2)  (8 x^2-6 x+3 ) \ln
    (1-\frac{x}{2} )+\ln (1-x)  (-32 x^6 \ln (4-2 x)\nonumber\\
&&+32 (2 x (x (2
   x-3)+2)-1) x^2 \ln (2-x)+x (x (x (4 x (x (4 x (5+2 \ln (2))-57\nonumber\\
&&-20 \ln
   (2))+66+50 \ln (2))-167-172 \ln (2))+7 (9+8 \ln (2)))-13-4 \ln (2))\nonumber\\
&&-2
   (x-1)^2 (2 x-1)^3 x \ln (x+1)+1 )-2 (x-1) x (x (4 x-1)+1) (1-2 x)^2 \ln
   (x) \ln (2 x+1) )\,,\nonumber
\end{eqnarray}
\begin{eqnarray}
f_{13}&=&  (x  (6  (x  (x  (72 x^2-62
   x-23 )+41 )-13 ) (2 x-1) \ln (2)+8 (47 x-20) (2 x-1)^3-3 \pi ^2
   (x (8 x (x (20 x-47)+39)\nonumber\\
&&-101)+8)+36 (x (2 x (2 x (2 x (x+3)-17)+29)-23)+4) \ln
   ^2(2) )+18  (16 x^2 (x-1)^4 \ln ^2(1-x)\nonumber\\
&&+x  ((1-2 x)^2  ((2
   x-1)  (-6  (x^2-1 ) \ln ^2 (\frac{x+1}{2} )+(-2 (x-2)
   x-3) \ln ^2(x)+2 (x-1)^2 \ln (2) \ln (x+1) )\nonumber\\
&&+2 (x ((5-4 x) x-2)+1) \ln
   (x) \ln (2 x+1) )-(x-2) (3 x-2)  (8 x^2-6 x+3 ) (2 x-1) \ln
    (1-\frac{x}{2} )\nonumber\\
&&+6 (x-2) x (2 x-1)^3 \ln
   ^2 (1-\frac{x}{2} ) )+(x-1) \ln (1-x)  (-32 x^2 (x-1)^3 \ln
   (2-x)\nonumber\\
&&-2 x (2 x-1)^3 (x-1) \ln (x+1)+x (x (4 x (x (20 x-37-20 \ln
(2))+29+30
   \ln (2))-51\nonumber\\
&&-52 \ln (2))+12+4\ln (2))-1 ) )-288 x^2 (x-1)^4 \ln (2)
   \ln (4-2 x) )\,,\nonumber
\end{eqnarray}
\begin{eqnarray}
f_{14}&=&(x (180 x-58 x \ln (2)-139+40 \ln (2))-2 ((x-1) x+2) \ln
(1-x)+2
   ((x-1) x+2) \ln (2-x)\nonumber\\
&&+64 (x-1) x \ln (2 x)-2-4 \ln (2))\,,\nonumber
\end{eqnarray}
\begin{eqnarray}
f_{15}&=&  (x  (x^2 \ln (128 x)-32 (x-1) x \ln (2-2 x)+(3-4 x)
   \ln (2 x)-(x-3) (x-1) \ln (x+1) )\nonumber\\
&&+x (x (x (15+22 \ln (2))-9-28 \ln
   (2))-9)+3 )\,,\nonumber
\end{eqnarray}
\begin{eqnarray}
f_{16}&=&  (13 \ln (2)  (\ln (32)-6 x^2 )+x (384 x-380+(19-117 \ln
   (2)) \ln (2))+20-30 \ln (2) )+5 \pi ^2 (12 x-7)\nonumber\\
&&+6 \ln ^2(x)-324 \ln
   ^2 (\frac{x+1}{2} )+6  (64 (x-1) x \ln ^2(1-x)+2 x  (-27 (x-2)
   \ln ^2 (1-\frac{x}{2} )+4 (9-8 x) \ln ^2(x)\nonumber\\
&&+27 x \ln
   ^2 (\frac{x+1}{2} ) )-2 ((x-1) x+2) \ln (2) \ln (2-x)+(2 x
   ((x-4) \ln (2)+16)\nonumber\\
&&+6 \ln (2)) \ln (x+1) )+192 \ln (x+1)\,,\nonumber
\end{eqnarray}
\begin{eqnarray}
f_{17}&=& (-6 (x-1) x \ln ^2(1-x)+6 (x-1) x \ln ^2(x)+2 (2 x-1)
 (\pi ^2-3 \ln
   ^2(2) )\nonumber\\
&&+24 (x-1) x \ln (2) \ln (1-x)-24 (x-1) x \ln (2) \ln (x)
)\,,\nonumber
\end{eqnarray}
\begin{eqnarray}
f_{18}&=&32 (x-1) x \text{Li}_2 (\frac{x-1}{x} )-32 (x-1) x
   \text{Li}_2 (\frac{x}{x-1} )-192 x^2-46 x^2 \ln ^2(1-x)+46 x^2 \ln ^2(x)\nonumber\\
&&+64 x^2 \ln
   (2-2 x) \ln (1-x)-72 x^2 \ln (2) \ln (1-x)-48 x^2 \ln (1-x)+72 x^2 \ln (2) \ln (x)-48 x^2
   \ln (x)\nonumber\\
&&-64 x^2 \ln (x) \ln (2 x)-10 \pi ^2 x+160 x+46 x \ln ^2(1-x)-46 x
\ln ^2(x)+28 x
   \ln ^2(2)-64 x \ln (2-2 x) \ln (1-x)\nonumber\\
&&+72 x \ln (2) \ln (1-x)+64 x \ln (1-x)-72 x \ln (2)
   \ln (x)+16 x \ln (x)+64 x \ln (x) \ln (2 x)\nonumber\\
&&+8 x \ln (2) \ln (2)-44 x \ln (2)-16 \ln
   (1-x)+4 \pi ^2-18 \ln ^2(2)+12\ln (2)+32 \ln (2)\,.\nonumber
\end{eqnarray}

\end{appendix}

\end{document}